\documentclass[12pt,preprint]{aastex}
\def\lap{\lower.5ex\hbox{$\; \buildrel < \over \sim \;$}}
\def\gap{\lower.5ex\hbox{$\; \buildrel > \over \sim \;$}}

\def\ergcm2s{${\rm erg\ cm^{-2}\ s^{-1}}$}

\def\ergscm2s{${\rm erg\ cm^{-2}\  s^{-1}}$}
\def\ergcms{${\rm erg\ cm^{-2}\  s^{-1}}$}
\def\cm-2{${\rm cm^{-2}}$}
\def\cms{${\rm cm^{-2}}$}
\def\ergs{${\rm erg\ s^{-1}}$}

\def\msol   {{M$_{\odot}$}}
\def\msolyear   {{M$_{\odot} {\rm yr}^{-1}$}}
\def\lax    {${_<\atop^{\sim}}$}

\def\mathfont#1{\ifmmode{#1}\else{$#1$}\fi} 
\def\lae{\mathrel{<\kern-1.0em\lower0.9ex\hbox{$\sim$}}}  
\def\gae{\mathrel{>\kern-1.0em\lower0.9ex\hbox{$\sim$}}}  
\def\kms{\ifmmode{{\rm km\ s}^{-1}}\else{${\rm km\ s}^{-1}$}\fi}

\def\msun{\ifmmode{\ {\rm M}_\odot}\else{$ {\rm M}_\odot$}\fi}  
\def\msunyr{\ifmmode{\msun \ {\rm yr}^{-1}}\else{$\msun \ {\rm yr}^{-1}$}\fi}

\def\ref#1{\noindent\hangindent=24.0pt\hangafter=1{#1}\par}
\def\la{\hbox{\rlap{$<$}\lower.5ex\hbox{$\sim$}\ }}
\def\ga{\hbox{\rlap{$>$}\lower.5ex\hbox{$\sim$}\ }}
\def\lap{\lower.5ex\hbox{$\; \buildrel < \over \sim \;$}}
\def\gap{\lower.5ex\hbox{$\; \buildrel > \over \sim \;$}}

\def\sgra   {Sgr~A$^*$}
\def\ms    {M31$^*$}

\def\msol   {{M$_{\odot}$}}

\def\kms    {~km~s$^{-1}$}

\def\ergscm2s  {~erg~cm$^{-2}$~s$^{-1}$}
\def\cm2s{~cm$^{-2}$~s$^{-1}$}
\newcommand{\cmsq}{~cm$^{-2}$}
\def\cm3{~cm$^{-3}$}
\def\nh    {N$_{\rm H}$}

\def\chandra	{{\it Chandra}}

\received{2009 xx xx}

\begin{document}

\shortauthors{Garcia et al.}
\shorttitle{M31* with Chandra}
\slugcomment{July 2009, for {\em The Astrophysical Journal}}

\title{X-ray and Radio Variability of M31*, The Andromeda Galaxy
  Nuclear Supermassive Black Hole}

%
%
\author{Michael R. Garcia\altaffilmark{1}, 
Richard Hextall\altaffilmark{1,2},
Frederick K. Baganoff\altaffilmark{3},
Jose Galache\altaffilmark{1},
Fulvio Melia\altaffilmark{4},
Stephen S. Murray\altaffilmark{1},
Frank A. Primini\altaffilmark{1}, 
Lor\'{a}nt O. Sjouwerman\altaffilmark{5},
and Ben Williams\altaffilmark{6}}

\altaffiltext{1}{Harvard-Smithsonian Center for Astrophysics, 60
Garden Street, Cambridge, MA 02138; garcia@head.cfa.harvard.edu}
\altaffiltext{2}{Department of Physics, University of Southampton, UK}
\altaffiltext{3}{Kavli Institute for Astrophysics and Space Research,
  Massachusetts Institute of Technology, Cambridge MA 02138}
\altaffiltext{4}{Physics Department,The Applied Math Program, and
  Steward Observatory, The University of Arizona, 
Tucson AZ 85721}
\altaffiltext{5}{National Radio Astronomy Observatory, Socorro NM
  87801}
\altaffiltext{6}{Department of Astronomy, University of Washington,
  Seattle WA 98195} 


\begin{abstract} 

We confirm our earlier tentative detection of \ms\ in X-rays and
measure its light-curve and spectrum.  Observations in 2004-2005 find
M31* rather quiescent in the X-ray and radio.  However, X-ray observations
in 2006-2007 and radio observations in 2002 show M31* to be highly
variable at times.  A separate variable X-ray source is found near P1,  
the brighter of the  two optical
nuclei.  The apparent angular Bondi radius of M31* is 
the largest of any black hole, and large enough to be well resolved
with Chandra.  The diffuse emission within this Bondi radius is found
to have an X-ray temperature $\sim 0.3$ keV and density 0.1
cm$^{-3}$, indistinguishable from the hot gas in the surrounding
regions of the bulge given the statistics allowed by the current
observations.   The X-ray source at the location of 
M31* is consistent with a point source and a power law
spectrum with energy slope $0.9 \pm 0.2$.   Our identification of this
X-ray source with M31* is based solely on positional coincidence.

\end{abstract}


\section{Introduction}

SMBHs in galactic nuclei spend the vast
majority of their life accreting at very low rates, but our
understanding of how this accretion occurs is relatively poor compared
to our understanding of what happens at high rates.  At high rates a
slim accretion disk often forms and $\sim$10\% of the accretion energy
is radiated.  At low rates the accretion becomes radiatively
inefficient, but we are uncertain what fraction of the accretion
energy is radiated, and what fraction of the gas accreted at large
radii actually reaches the black hole event horizon.

While it is clear that  some sort of a radiatively inefficient
accretion flow (RIAF) occurs, its form as a magnetically dominated
inflow \cite{shvartsman.1971, melia.1992}, 
an ADAF (Advection Dominated Accretion
Flow; \cite{narayan.yi.94}), CDAF (Convective Dominated Accretion Flow
in which advection occurs but is moderated by convection;
\cite{cdaf.Ig, cdaf.narayan, cdaf.quat}), an ADAF \& wind (in which a
large fraction of the accreted material is blow out in a wind before
reaching the SMBH; \cite{narayan.wind.95}); sometimes called an ADIOS
(Advective Dominated Inflow Outflow Solution; \cite{bb.adios,
hb.adios}), or something else is unclear. One of the most promising ways
to determine which of these alternatives is correct is to image the
accretion flow on a sub-Bondi scale and therefore determine the
structure of the flow.  Spectra taken on sub-Bondi scale would also be
highly discriminating.

    Perhaps the best candidate for such a study is the
 SMBH in M31, called hereafter \ms.  The
mass of \ms\ has previously been estimated as $3 \times 10^7$~\msol
\citep{kormendy.bender.99}, but more recent HST spectroscopy indicates the
mass is $1.4 ^{+0.7}_{-0.3} \times 10^8$~\msol \citep{Bender.2005}.
This higher mass means the Bondi radius is not the $0.9''$ previously
estimated \citep{garcia.2005} but $\sim 5''$ (see below), making it
the largest of any known SMBH (see Figure~1).  Within the Bondi radius
Chandra detects several point sources and also diffuse gas which is
present throughout the bulge of M31 \citep{shirey.xmm, taka.m31,
li.wang.2007}.  Our earlier work \citep{garcia.m31*.2000} suggested
that a super-soft point source near the nucleus was the X-ray counterpart
of M31*, but subsequent alignment with M31 globular clusters
\citep{barmby.huchra.2001} revealed that this was not the case
\citep{garcia.m31*not.2001, garcia.2005}.  \chandra\ HRC imaging
revealed that the X-ray source closest to the nucleus is made up of
two partially resolved components, one of which is within
$\sim 0.1''$\ of the position of P3 and which we tentatively identified
as the  $\sim
10^{36}$ \ergs\/ X-ray counterpart of \ms\ \citep{garcia.2005}.
In this work we confirm the detection
and measure the variability of this SMBH.

Hubble imaging revealed an unusual double nucleus within M31
\citep{lauer.93} which has been successfully modeled as an eccentric
torus of stars viewed nearly edge on \citep{tremaine.1995}.  The
optically brighter end of the torus is known as P1, the optically
fainter (but UV brighter) end as P2.  More recent Hubble imaging and
spectroscopy revealed that embedded within P2 is a dense cluster of
several hundred A-stars which are very UV bright. This cluster has
been dubbed P3 and its brightness peaks at essentially the same
position as that of P2 \citep{Bender.2005}.  HST spectroscopy reveals a dark
mass (presumed black hole) located at the center of this cluster with
an estimated mass of $1.4 ^{+0.7} _{-0.3} \times 10^8$~\msol
\citep{Bender.2005}.  The presence of so many young A stars in the
nucleus of M31, which otherwise is made up of an old stellar
population, is a mystery.  The same phenomenon is also seen near \sgra\
\citep{ghez.youth.2003}.  \cite{demarque.virani} suggest that these
apparently young stars may actually be old and relics of stellar
collisions in the very dense nuclear regions, but recent HST spectra
indicate that at least in the case of the Galactic center these are
genuine young, massive stars \citep{ghez.2003,martins.2008}. 
A suggested source for the gas which later collapses to form 
the P3 star cluster is stellar mass loss from the older stars in the
orbiting torus \citep{chang.2007}.

Spitzer imaging reveals that M31 looks less like a spiral galaxy than
previously thought.  Instead, the warm gas and dust is in the form of
a set of rings, apparently due to transits of M32  through the 
center of the plane of
the M31 disk \citep{barmby.2006, gordon.2006}.  The most
recent of these transits took place 210 million years ago
\citep{block.2006}, and would have triggered a burst of star formation.

The association of the radio source ($\sim 30$ $\mu$Jy at 3.6 cm) with
a nuclear black hole by \cite{crane.93} and \cite{melia92} and
subsequently by all others so far is supported by the
detection of an unresolved ($< 1$ pc or $0.35''$) radio source
at the position of the stellar nucleus (within $\sim 1/2''$,
\cite{garcia.2005}), and radio variability which is comparable to the
radio variability of \sgra\ \citep{lorant.2005}.

While \ms\ is 100x further away than the SMBH in the center of our
Galaxy, it suffers much less reddening: $A_V \sim 1$
\citep{garcia.m31*.2000}, whereas $A_V \sim 30$ for \sgra . The
dominant temperature of the diffuse gas in the core of M31 is $\sim
0.3$ keV (\cite{anil.2001, shirey.2001, li.wang.diffuse,
bogdan.gilfanov.diffuse}) as compared to 1.0 keV \citep{baganoff.2003}
around \sgra .  A comparative study of these
 two closest SMBH (\ms\ and \sgra) will deepen our understanding of
 the physics of black hole accretion and emission mechanisms at low
 rates. In this regard, \ms\ and \sgra\ offer interesting differences
 in that \ms\ is 3x less luminous in the radio but 1 to 3 orders of
 magnitude more luminous in the X-ray.

\begin{figure}
\plotone{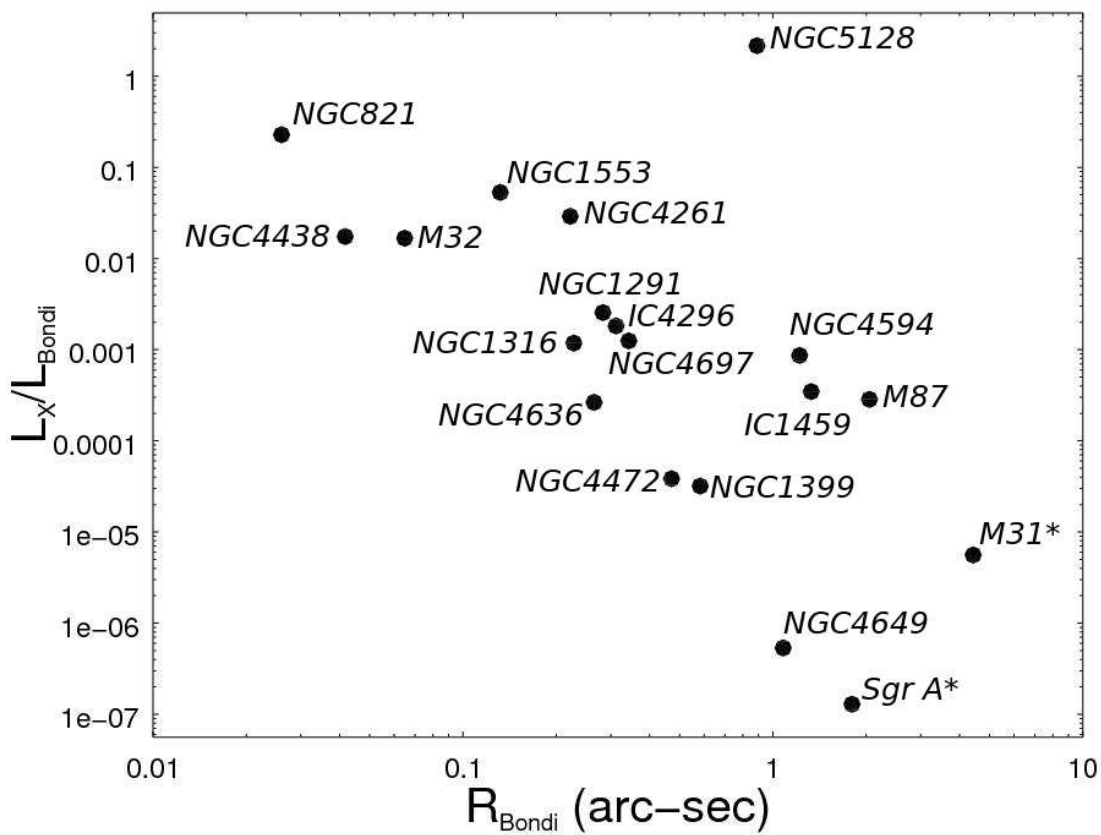}
\caption{
The Bondi radii of nearby SMBH vs. their apparent X-ray 
luminosity (or upper limits) in units of the expected Bondi luminosity. 
Objects in the lower right of this diagram provide simultaneously secure and 
severe constraints on accretion and emission models for these SMBH. 
In this regard, \ms\ is the outstanding object. }
\end{figure}

\section{Observations}

M31 has been observed extensively with both \chandra\ 
(i.e., \cite{garcia.m31*.2000, kong.2003, williams.2006,
  distefano.2004}) and XMM-Newton 
(i.e., \cite{shirey.2001, pietsch.2007}).  Many of the \chandra\ observations
have been in the form of monthly short ($\sim$5~ks) snap-shots in order to
monitor the variability of the point source population, but there also
have been longer observations to search for rapid variability
\citep{kaaret.hrc} and to study source populations
\citep{distefano.2004, pietsch.2007}.  
The total \chandra\ exposure on M31 is now
$\sim 1.3$Ms and it continues to climb.  If one limits the selection
of data to only those taken with ACIS-I within 1 arcmin of the nucleus,
the total exposure is currently $\sim 250$~ks.  While \chandra\ is able
to view M31 10 months out of every year, visibility from XMM-Newton is
restricted to a few week long window every 6 months, so the XMM-Newton 
observations have taken the form of less frequent, but longer,
exposures \citep{trudy.xmm}.

The \chandra\ observations discovered the first
resolved SNR in an external galaxy \citep{kong.snr1}, and summed Chandra
observations have
provided a fiducial point for X-ray color-color diagrams which help to
separate unresolved X-ray source populations in more distant galaxies
into SNR, high and low mass X-ray binaries, and background AGN
\citep{kong.m31.l1,andyp.cc}.  These data-sets have also revealed
extensive structure in the diffuse emission
\citep{li.wang.diffuse,bogdan.gilfanov.diffuse}.  Multi-wavelength
studies of the \ms\ environment suggests the presence of an outflow
of hot gas from the nuclear region \cite{li.2009}.

Below we discuss the data-sets analyzed in this paper; first those taken with
the \chandra\ HRC and then those taken with the \chandra\ ACIS.

\subsection{Chandra HRC Observations}

In 2004/2005 we undertook a series of 4 moderate (50ks) HRC-I
exposures separated by ~1 month in order to confirm our earlier
possible detection of \ms\ and to search for any variability.  Two of
these exposures were subdivided into approximately equal sections and
separated by the normal 10 hour pause in observations while Chandra
transits the Earth's radiation belts.  We also include in our
analysis archival HRC observations made during 2006/2007 which were
designed to monitor optical nova \citep{henze.2009}, and a single 50~ks
observation from 2001 \citep{kaaret.hrc}. 
Table 1 lists the data-set we have analyzed, giving observation
 dates and exposure times.

\begin{table}[!ht]
{\centering
\begin{tabular}{| l | l | l | }
\hline
Obs-id & Start Date  & Exposure Time   \\
       & (mm/dd/yy)  &  (ks)  \\  \hline
1912 & 11/01/01 &  50.0 \\ \hline
5925 & 12/06/04 &  46.7  \\ \hline
6177 & 12/27/04 &  20.2  \\ \hline
5926 & 12/27/04 &  28.5  \\ \hline
6202 & 01/28/05 &  18.2  \\ \hline
5927 & 01/28/05 &  27.2  \\ \hline
5928 & 02/21/05 &  45.2  \\ \hline
7283 & 06/05/06 & 20.1  \\ \hline
7284 & 09/30/06 & 20.2 \\ \hline
7285 & 11/13/06 & 18.7 \\ \hline
7286 & 03/11/07 & 19.1 \\ \hline
8526 & 11/07/07 & 20.2 \\ \hline
8527 & 11/17/07 & 20.2 \\ \hline
8528 & 11/28/07 & 20.2 \\ \hline
8529 & 12/07/07 & 19.1 \\ \hline
8530 & 12/17/07 & 20.1 \\ \hline
\hline
\end{tabular}
\caption{This table shows the Observation ID's, Dates
   and Exposure Times for the 16 \chandra\ HRC images discussed
in this paper.  }}
\end{table}

In order to identify \ms\ within the \chandra\ images and then extract a
light-curve, it was first necessary to accurately determine the
location of \ms\ within the HRC images.  The absolute astrometric
accuracy of HST and \chandra\ images is \lax$1''$, which is insufficient
to uniquely identify sources in the crowded core of M31.  Within
narrow fields (\lax$1'$) it is possible to register HST and Chandra
images to $\sim 0.1''$ accuracy \citep{edmonds.2003}.  Over larger
fields the PSF of the \chandra\ mirror increases substantially, and this
along with the calibration of the HRC may limit the astrometric
accuracy even with registration \citep{aldcroft.2004}.

There are very few sources in common between our \chandra\ images and
the HST/ACS (F435W) images which show the M31 double nucleus, so we used a 2
step process to register our images.  As part of our search for
counterparts to black hole X-ray transients \citep{williams.2005} we
obtained several HST/ACS images of the M31 nucleus, and one of these
was first registered to the Local Group Survey \citep{lgs} image of
M31.  The second step is to register the \chandra\ images to the same
LGS image.  We allow for an x and y translation, a single change in
scale factor, and roll, but typically the latter two of these are
negligible while the translation is significant and $<1''$.

The HST/ACS to LGS registration typically uses 20 stars and has a
RMS of $0.03''$, and the registration of individual Chandra/HRC
images to the LGS uses 6 or 7 globular clusters and has a rms
of $ 0.1'' - 0.2''$.  One of these 7 globular clusters sometimes
falls below the \chandra\ detection threshold, limiting us to 6
occasionally.  We note that there are 9 globular clusters which are
in both the \chandra\ and LGS images, but 2 of these are between $6'$
and $10'$ from the nucleus where the PSF due to the \chandra\ mirror has
widened substantially. Including these distant clusters increases the
error in the registration to $\sim 0.3''$, so we have excluded
them. The remaining clusters are between $1'$ and $5'$ from the
nucleus. 
Clearly the \chandra\ to LGS registration error dominates the final
registration, and this error itself is dominated by the counting
statistics within the PSF (and associated registration errors) in the
individual globular clusters within the \chandra\ images.

In order to ensure the most accurate, and perhaps more importantly,
stable, registration, we summed the HRC data into a single image and
registered that image to the LGS images.  The summing process (CIAO
{\it reproject\_aspect}) uses the positions of bright X-ray sources to
register the individual HRC images to a common frame to better than
$0.05''$.  The accuracy of the registration of the summed Chandra
image to the LGS image as determined with IRAF {\it ccmap} is $0.1''$
rms.  We also summed the 2001/2004/2005 and 2006/2007 HRC data separately
and registered these two images to the LGS, again finding a rms
accuracy of $0.1''$.  Comparing these two (now registered) HRC images
to each other, we find an offset between bright sources of $0.2''$,
larger than the formal error in the rms.  Given that we are solving
for 4 parameters (x and y shift, scale factor, and roll) with 7
points, the resulting formal rms might not be expected to be Gaussian
- we therefore take $0.2''$ as our de-facto registration error between
the HST and \chandra\ images.

\begin{figure}[h]
\plottwo{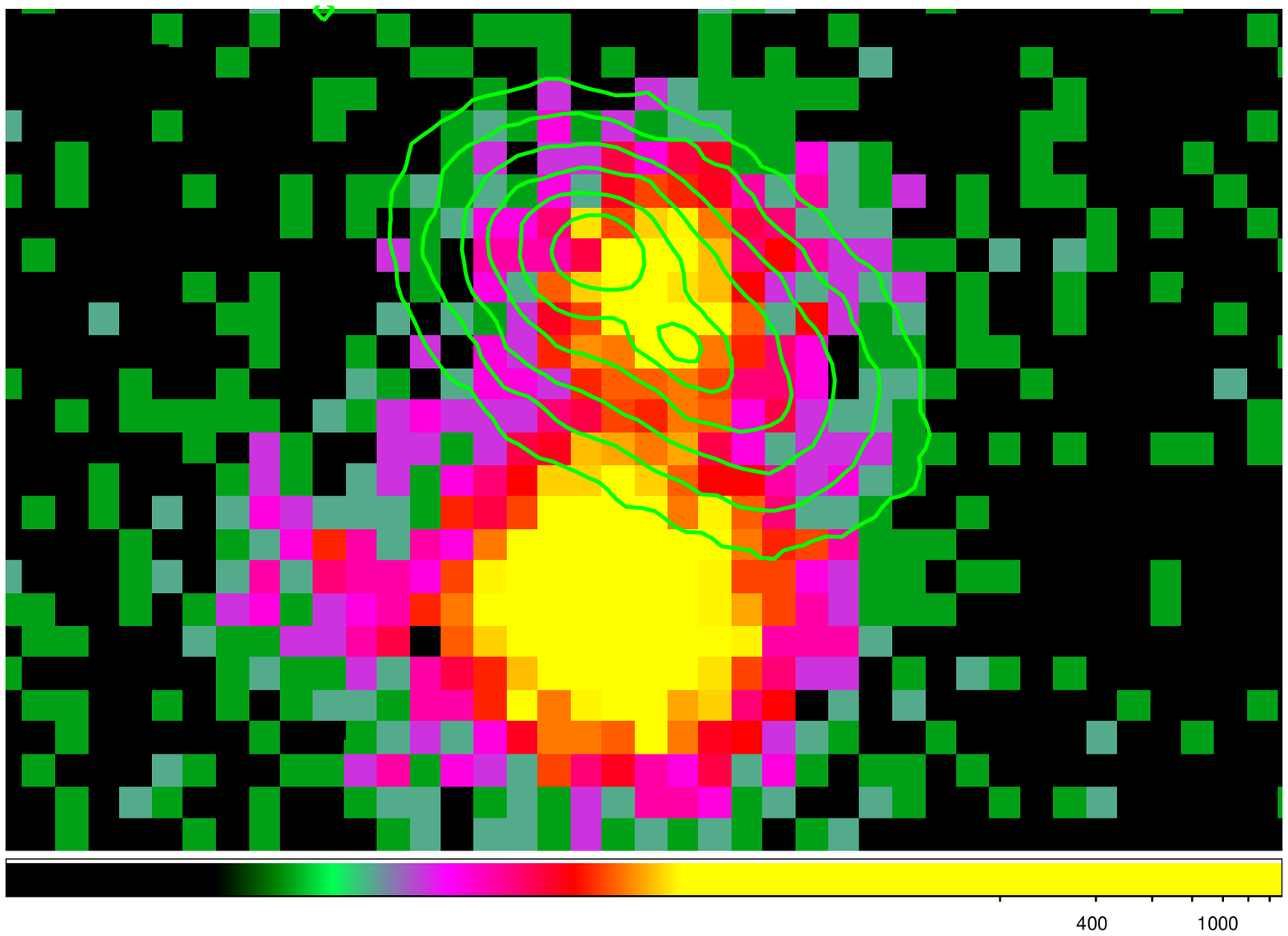}{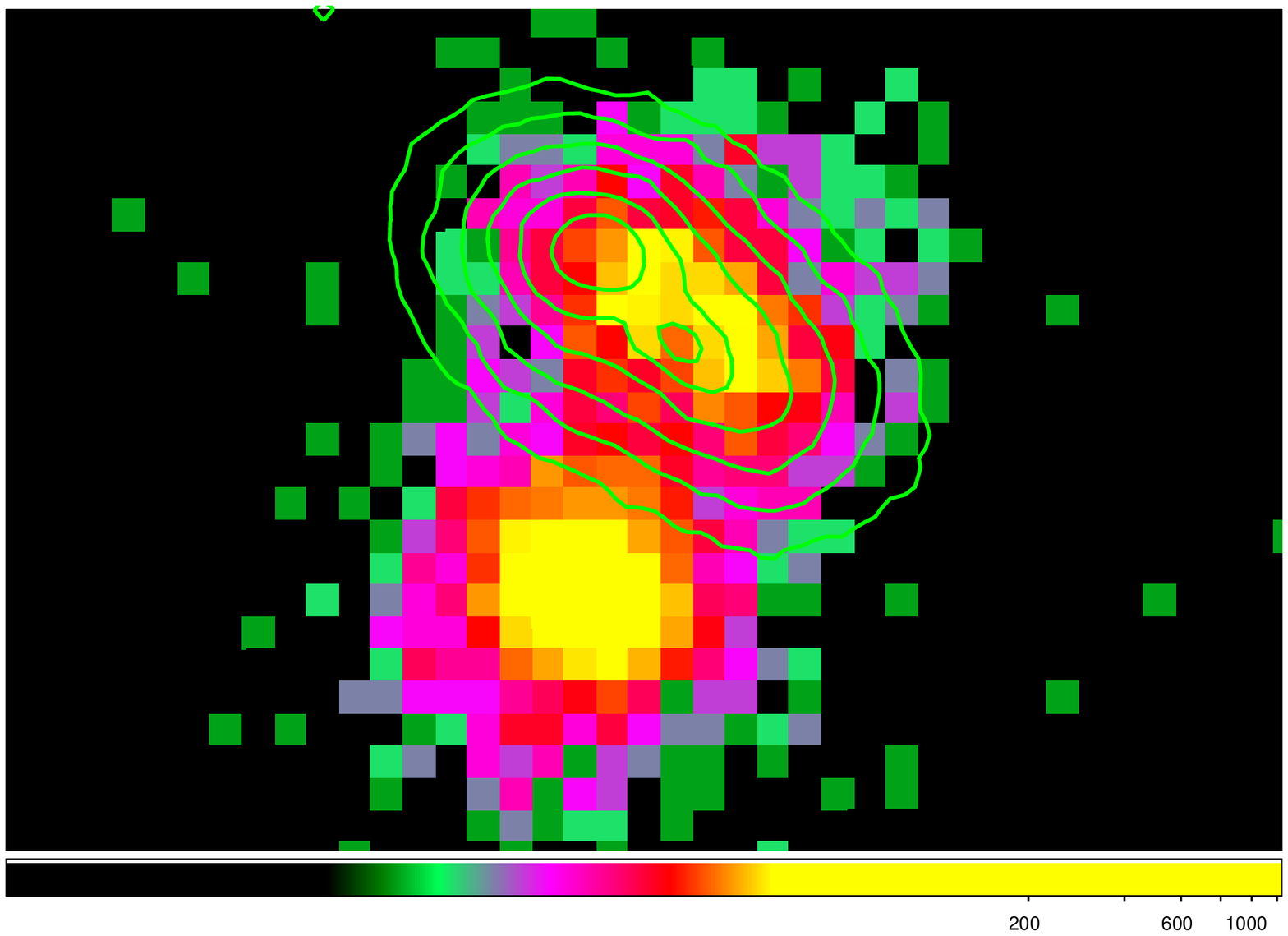}
\caption{The summed 2001/2004/2005 (left) and 2006/2007 (right) 
HRC images of the M31 
nucleus, registered to a common astrometric frame. The registration is
accurate to $\sim 0.2''$.  
The contours are from an HST/ACS image.  The
innermost contour levels are closed at the positions of 
P1 (upper left,or North-East) 
and P3 (lower right) and are separated by $0.5''$.  
In the 2004/2005 image the source associated with P3
(=M31*) is faint, while in the 2006/2007 image this source is stronger
and clearly separate from the source near P1. 
The super-soft source CXO J004244.2+411608
\citep{garcia.m31*.2000} is the bright source directly South of P1/P3. 
}
\end{figure}

These two images are shown in Figure~2.  The 2006/2007 image shows two
separate X-ray sources at the approximate locations of P1 and P3.  A
simple translation of $0.25''$ brings the Southwestern source to the
position of P3, and the Northeastern source $0.1''$ to the South of
P1.  Given the concurrence between the separation of the X-ray
sources, the angle between them, and P1 and P3 as seen in the HST
images, we identify the X-ray sources with M31* (=P3) and a source
within P1.  This concurrence can also be seen in Figure~3, which shows
the HST/ACS image with the HRC contours overlaid.  However, we cannot
entirely exclude the possibility that one or more transient sources
unrelated to P1 or P3 are responsible for the X-ray emission.

\begin{figure}[h]
\plottwo{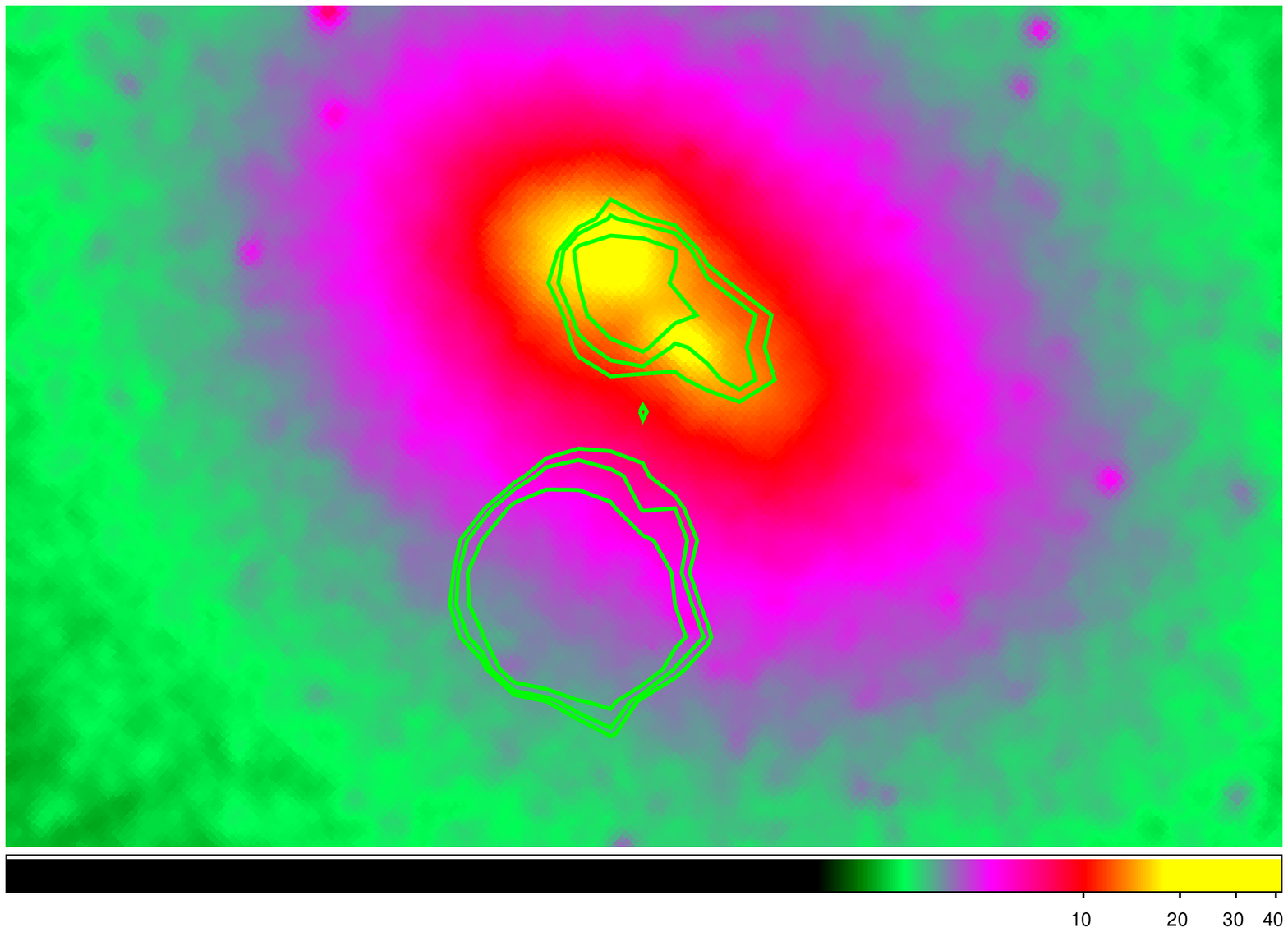}{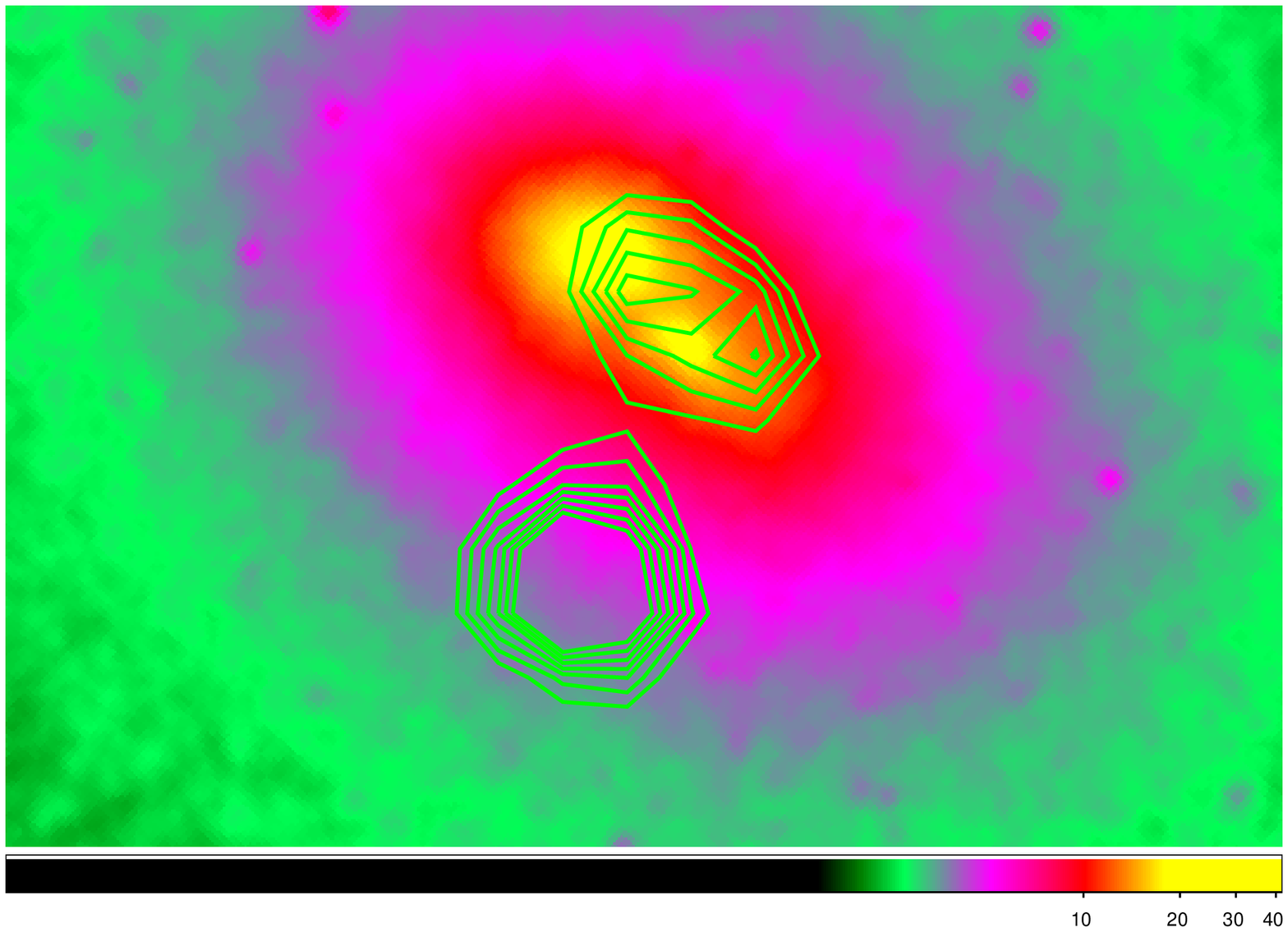}
\caption{A false color representation of the HST/ACS F435W image of
the M31 nucleus, with the contours from the HRC images overlaid.  The
left image shows the contours from the summed 430ks HRC data.  The
right image shows the contours from the 2006/2007 HRC data alone. 
A small East-West translation will bring the X-ray peaks on top of P3
and within $0.1''$ of P1. 
}
\end{figure}

Since counting statistics of the globular clusters in the individual
$\sim 20$~ks X-ray images limit the accuracy of the registration, we
rely on the registration of the merged 430~ks image and measure offsets
to P1 and P3 from the nearby super-soft source (CXO J004244.2+411608, 
\cite{garcia.m31*.2000}; hereafter SSS).  The
SSS has between 100 and 300 counts in each of the images, so can be
centroided to $\sim PSF/\sqrt(counts) \sim 0.05''$ in each of the
individual images.

In order to extract the HRC light-curves we centered extraction circles
at the locations of P1 and P3 as determined by the offsets from the
SSS (after the translation of $0.25''$ discussed above) 
on each of the images and counted the photons
therein.  We used radii of $0.4''$ centered at P1 and P3, as
shown in Figure~4.  P1 and P3 are sufficiently close that these
circles overlap, so we excluded the counts in the overlapping region.

\begin{figure}[h]
%
\plottwo{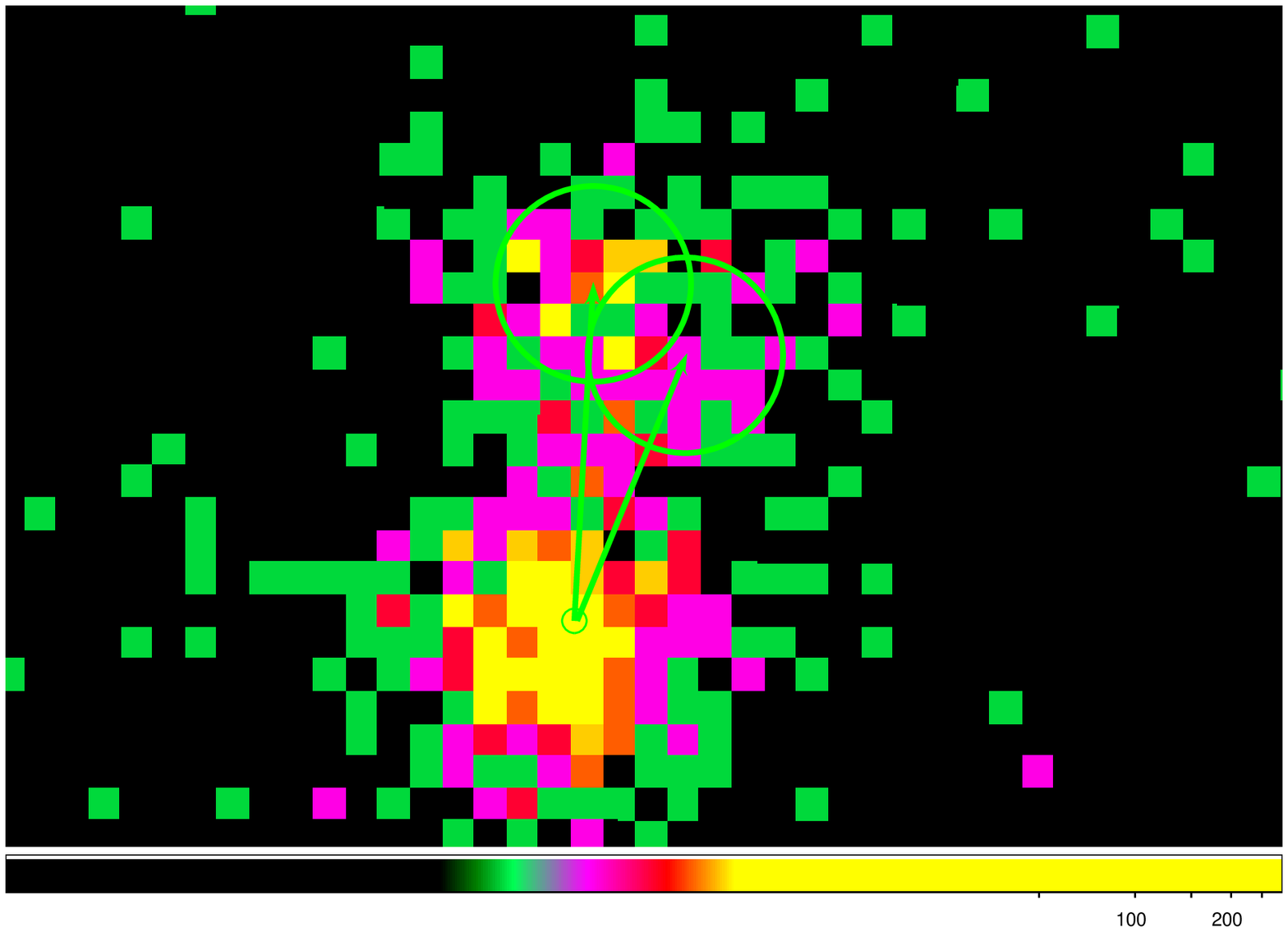}{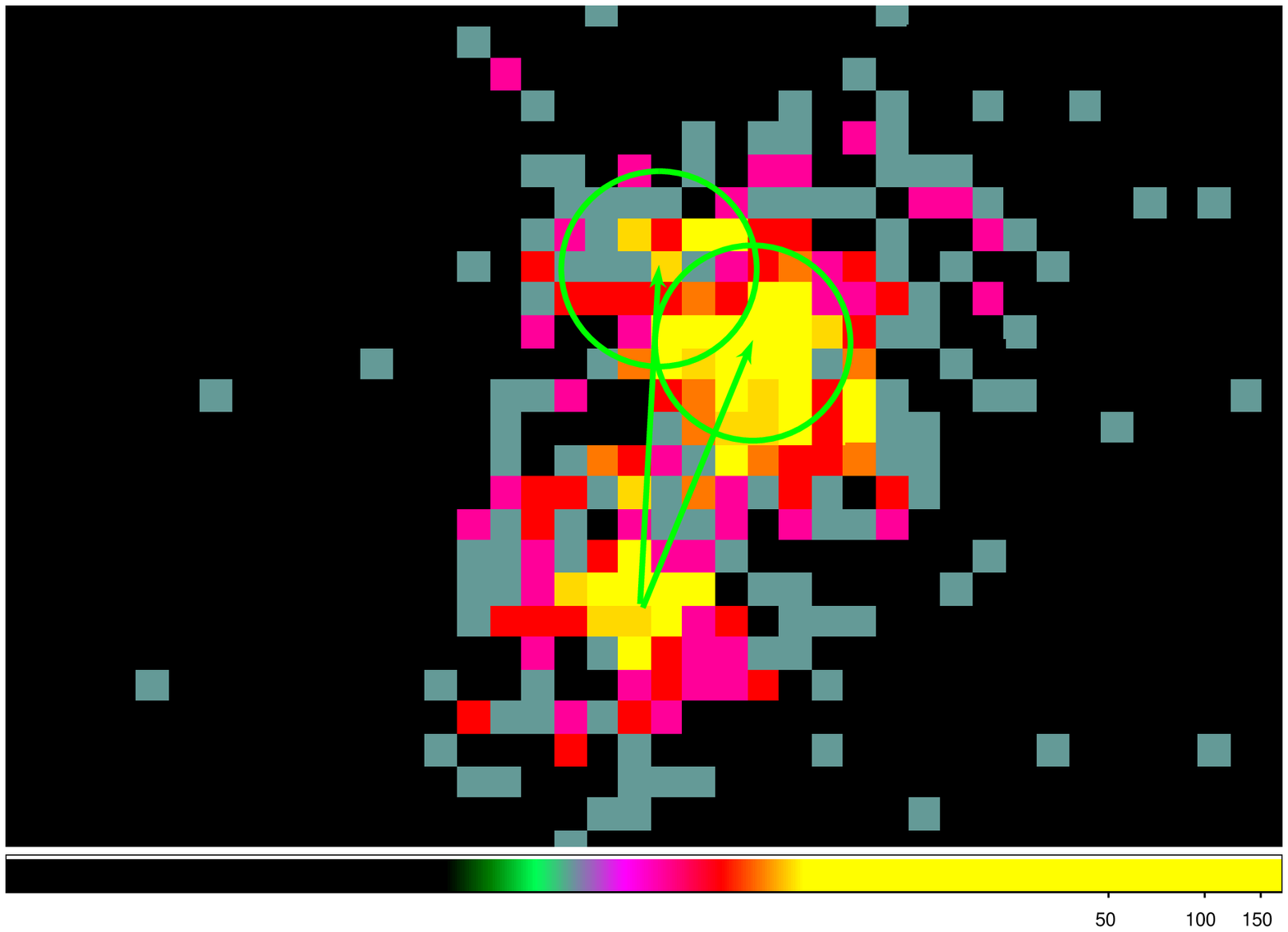}
\caption{Two of the 15 HRC images used in our light-curves.  
The left image from 2004 Dec 6.75 shows the source at P1 somewhat brighter than
that at P3, while the right image from 2008 Dec 7 shows P3 at its
brightest.  The $0.4''$ circles used to measure the counts are
re-centered for each observation by using an identical set of vectors
positioned to originate at the centroid of 
the super-soft source CXO J004244.2+411608 \citep{garcia.m31*.2000}
seen at the bottom of the images. 
Light curves were
generated from the counts in these $0.4''$ circles, excluding counts
from the overlapping region.  
}
\end{figure}

Figure~5 shows the resulting light-curves. 
There is clearly some correlation between the P1 and P3 light-curves,
as is expected due to the proximity of the two sources.  
The $0.4''$ extraction circle contains 50\% of the flux from a point
source and the two sources are only $0.5''$ apart.

The most dramatic variability we find is in December 2007 (day 519 to
529), where M31* varies by a factor of 3 in 10 days. The maximum
variability is between day 117 and 519, where we see a factor of more
than 10 variation.  In order to search for variability on short
timescales, we divided the day 519 observation (where \ms\ was
brightest) into 5 approximately equal 4000~s intervals.  
Figure~6 shows the resulting light-curve.  
While there are apparent $\sim 2\sigma$
deviations, a $\chi^2$ test to the hypothesis of a constant source has 
a  6\% chance of being satisfied, so significant variability cannot be
convincingly argued. 

In order to test the sensitivity of the light-curve to possible errors
in the registration, we shifted the extraction circles $0.05''$ both RA
and DEC, as this is the typical error in measuring the centroid of the
SSS which is used as the reference point.  While the overall character
of the resulting light-curve remains the same, some of the points do
see $>2\sigma$ changes (see Figures 5 and 6).  

Assuming the spectral
shape found with ACIS (see below) of $\alpha_\nu = 0.9$ and 
${\rm N_H} = 6 \times 10^{20}$ \cmsq , 
a measured rate of 1.0 c/ks
 corresponds to an emitted luminosity over $0.3 - 7.0$
keV of $2.25 \times 10^{36}$ \ergs\ in the full HRC beam, 
or $3.0 \times 10^{36}$ \ergs\ over $0.1 - 7.0$ keV.  
M31* therefore ranges in
luminosity from $0.6 - 20 \times 10^{36}$ \ergs\ ($0.1 - 7.0$ keV).
We note that if the source we associate with \ms\ is in fact 
unrelated, then these number are upper limits and the X-ray luminosity
of \ms\ must at times be below $6 \times 10^{35}$ \ergs .


\begin{figure}[h]
\plottwo{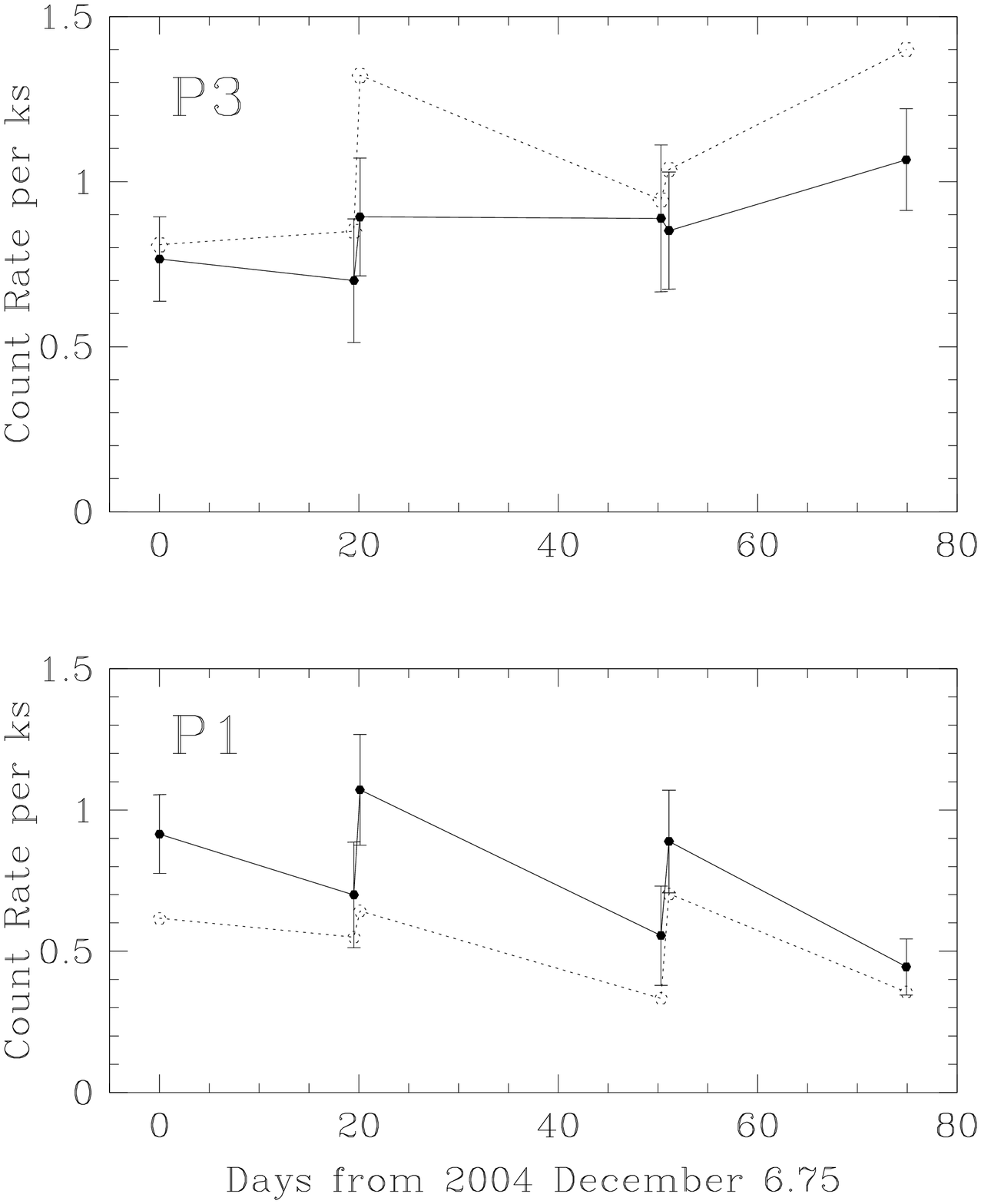}{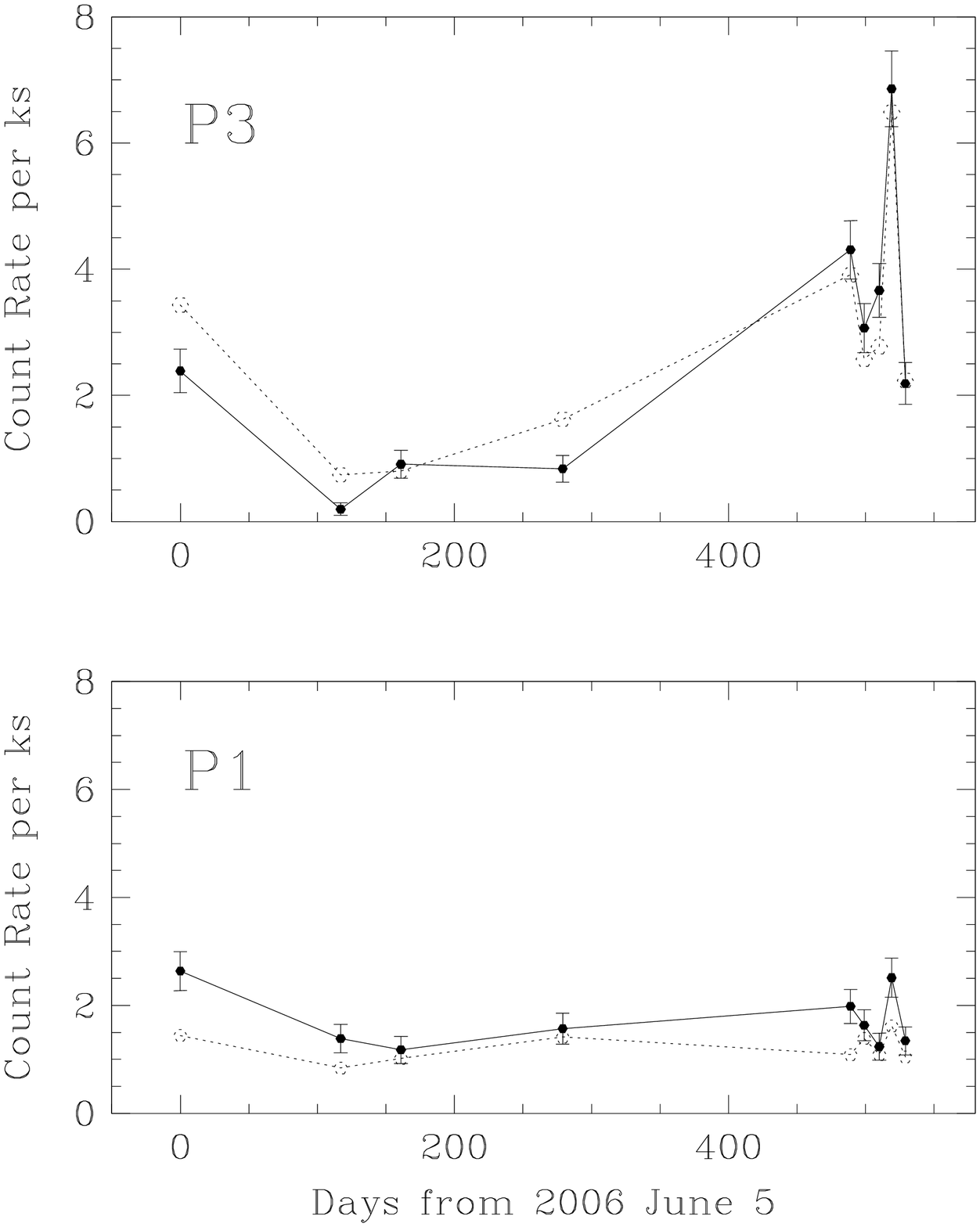}
\caption{
Light-curves for P3 = M31* and the source near P1 
as extracted from our 2004/2005 HRC data (left) and archival 2006/2007 (right)
HRC data.  A counting rate of 1.0 counts per ks corresponds to 
a luminosity of $3 \times 10^{36}$ \ergs\ ($0.1-7.0$ keV) 
at the 780~kpc distance of M31. Error-bars are $1\sigma$ counting
statistics only.  The dashed line indicates the sensitivity of the
light-curve to $1\sigma$ shifts in the registration and therefore count
extraction procedure.  We 
do not include the 2001 observation in the light-curve due to its 3 year
offset. 
\label{p3.lightcurve}}
\end{figure}

\begin{figure}[h]
\includegraphics[scale=0.4]{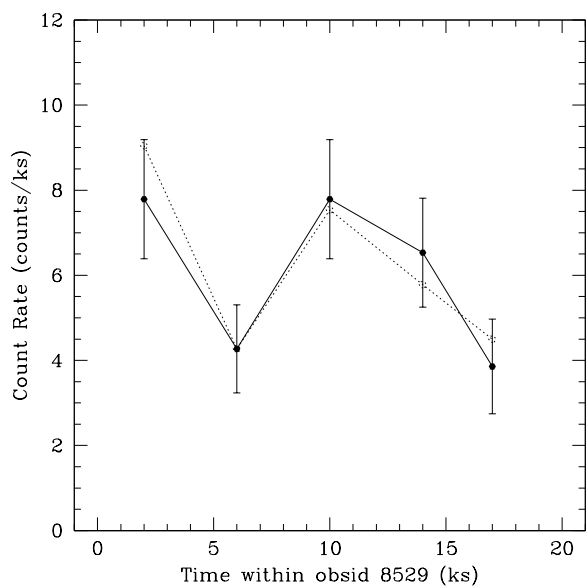}
\caption{
Light-curve for P3 = M31* during obsid 8529 (day 529 in Figure~5).
The data were divided into 4000 sec intervals in order to measure the
rate. While there are $\sim 2\sigma$ deviations, a $\chi^2$ test to a
constant source finds $\chi^2/\nu = 8.87/4$, which has a 6\% chance of
originating from a constant source. Short time scale variability is
therefore not detected.   As before, the dashed line and points
represent the light-curve extracted from slightly shifted positions.
\label{o8529.lightcurve}}
\end{figure}

\clearpage

\subsection{Chandra ACIS}

While the HRC data gives excellent spatial resolution and allows us to
determine an accurate light-curve for \ms , They do  not provide
sufficient energy resolution to allow us to search for spectral
signatures of the Bondi flow/ADAF.  We therefore investigate the
archival ACIS-I and ACIS-S data.

\begin{table}[!ht]
\centering
\begin{tabular}{| l | l | l |}
\hline
Obs-id & Start Date & Exposure Time  \\
       & (mm/dd/yy) &  (ks) \\ \hline
303 & 10/13/99 & 11.8 \\ \hline
305 & 12/11/99 & 4.1 \\ \hline
306 & 12/27/99 & 4.1 \\ \hline
307 & 01/29/00 & 4.1 \\ \hline
308 & 02/16/00 & 4 \\ \hline
311 & 07/29/00 & 4.9 \\ \hline
312 & 08/27/00 & 4.7 \\ \hline
1581 & 12/13/00 & 4.4 \\ \hline
1582 & 02/18/01 & 4.3 \\ \hline
1583 & 06/10/01 & 4.9 \\ \hline
4360 & 08/11/02 & 4.9 \\ \hline
4678 & 11/09/03 & 3.9 \\ \hline
4679 & 11/26/03 & 3.8 \\ \hline
4680 & 12/27/03 & 4.2 \\ \hline
4681 & 01/31/04 & 4.2 \\ \hline
4682 & 05/23/04 & 3.9 \\ \hline
4719 & 07/17/04 & 4.1 \\ \hline
4720 & 09/02/04 & 4.1 \\ \hline
4721 & 10/04/04 & 4.1 \\ \hline
4722 & 10/31/04 & 3.8 \\ \hline
7064 & 12/04/06 & 23.2 \\ \hline
7136 & 01/06/06 & 4 \\ \hline
7137 & 05/26/06 & 3.9 \\ \hline
7138 & 06/09/06 & 4.1 \\ \hline
7139 & 07/31/06 & 4 \\ \hline 
7140 & 09/24/06 & 4.1 \\ \hline
8183 & 01/14/07 & 4 \\ \hline 
8184 & 02/14/07 & 4.1 \\ \hline
8185 & 03/10/07 & 4 \\ \hline
\end{tabular}
\caption{This table shows the Observation ID's, Dates, Start
  Times and Exposure Times for the 29 ACIS-I images merged together and 
shown in Figure 7 (left).  Total exposure time is 148~ks.}
\end{table}

We used 29 ACIS-I exposures aimed within $1'$ of \ms\ which total to
148ks of exposure time.  This summed exposure is shown in Figure~7
(left) as a smoothed 3-color X-ray image where the soft (0.3-1.0 keV),
medium (1.0-2.5 keV), and hard (2.5-7.0 keV) bands are color coded as
red, green, and blue respectively.  The $5''$ Bondi radius is shown
here in green.  The emission due to hot diffuse gas is seen throughout
this image and generates $\sim 1$ event per pixel in this image.
Note that we do not include the ACIS-S observation discussed below,
obsid 1575, in this sum due to the substantially different energy
response of ACIS-S vs. ACIS-I.  

We searched for an enhancement in the flux at (or inside) of the Bondi
radius by generating radial profiles of the flux, centered on \ms\ ,
in several wedges that are free of point sources.  Defining the ray to
the North as zero degrees and moving counterclockwise, we generated
these profiles over the azimuth ranges of 40 to 70 degrees, 100 to 150
degrees, and 245 to 280 degrees.  In the first two of these azimuth
ranges we found a slight excess of counts near the Bondi radius (above
the local background), amounting to 10 counts from $3''$ to $5''$ in
the first region and 20 counts from $4''$ to $6''$ in the second
region.  We then used MARX to estimate the number of counts that would
be scattered into these regions from the surrounding point sources,
and found that this could account for $\sim 1/3$ of the observed
excess.  Note that there is clear spatial structure in the diffuse
emission at larger radii ($10'' - 20''$, see \cite{li.2009}) which,
if present at or within the Bondi radius, could easily account for the
apparent excess of 10 to 20 counts seen in two of our regions.
Assuming that the results in these three azimuth ranges are typical of
what would be found for the diffuse emission at all azimuth ranges,
and given that these three ranges sum to $\sim 1/3$ of the total, we
take the apparent excess of 30 total counts to correspond to a limit
three times higher, or $\sim 100$ counts, on any truly diffuse
emission due to the Bondi flow.  This limit corresponds to an emitted
flux of less than $9 \times 10^{-15}$~\ergcms (0.3-7.0 keV) assuming a
temperature of 0.3 keV, or a luminosity at 780 kpc of $< 6 \times
10^{35}$~\ergs .

Given that we do not detect a significant enhancement in the flux at
(or inside) the Bondi radius, we would not expect to detect a
temperature change either as any such change would need to be
co-incidentally offset by a change in electron density.  None the
less, we did select energy bands of Figure~7 such that approximately
equal counts are produced in the soft and medium bands in order that
any change in the $\sim 0.3$ keV temperature of the diffuse gas would
be emphasized.  As evidenced by the lack of a color change in
Figure~7, there is no apparent change in the temperature of the
diffuse gas at the Bondi radius.

\begin{figure}[h]
\plottwo{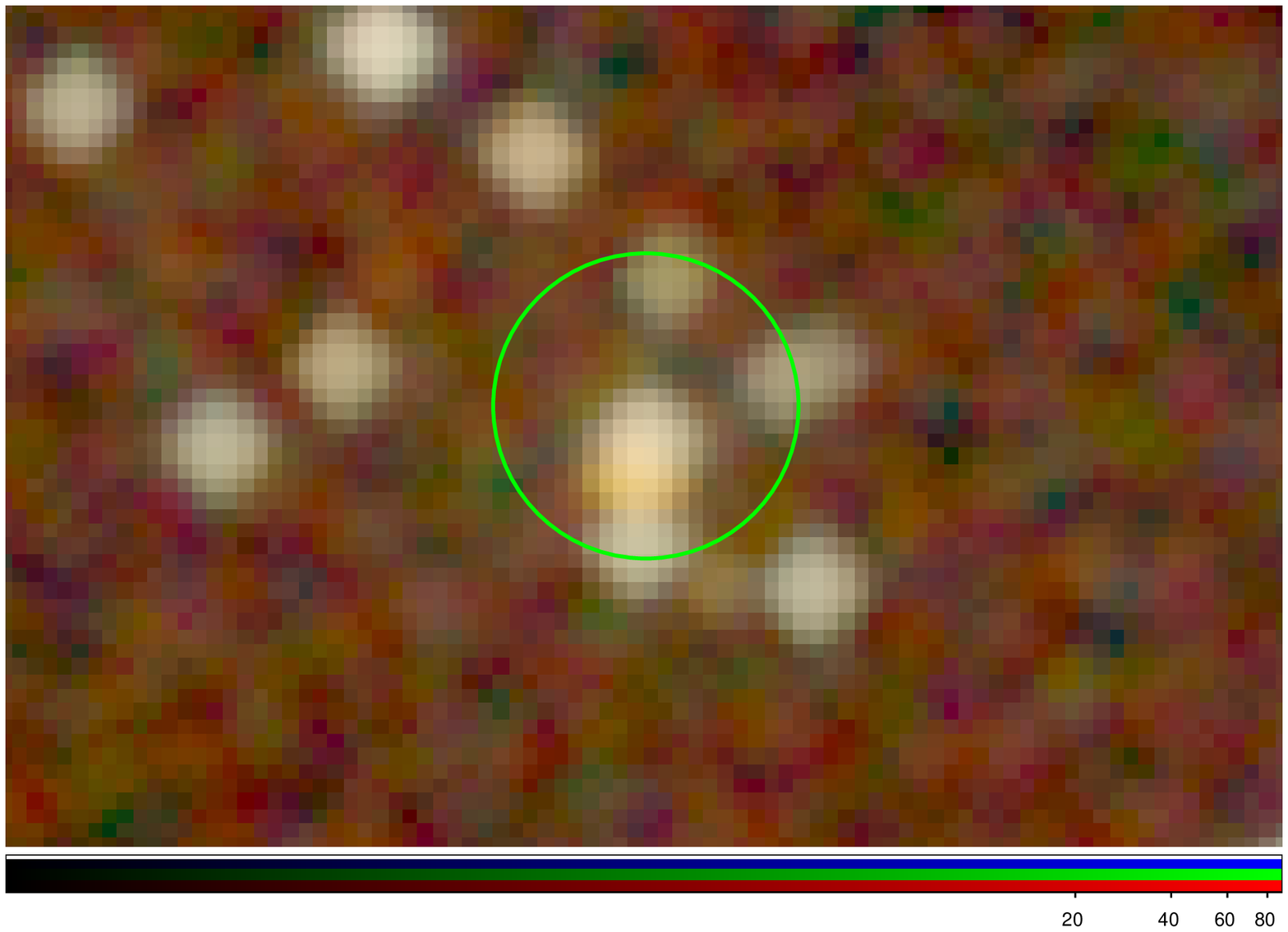}{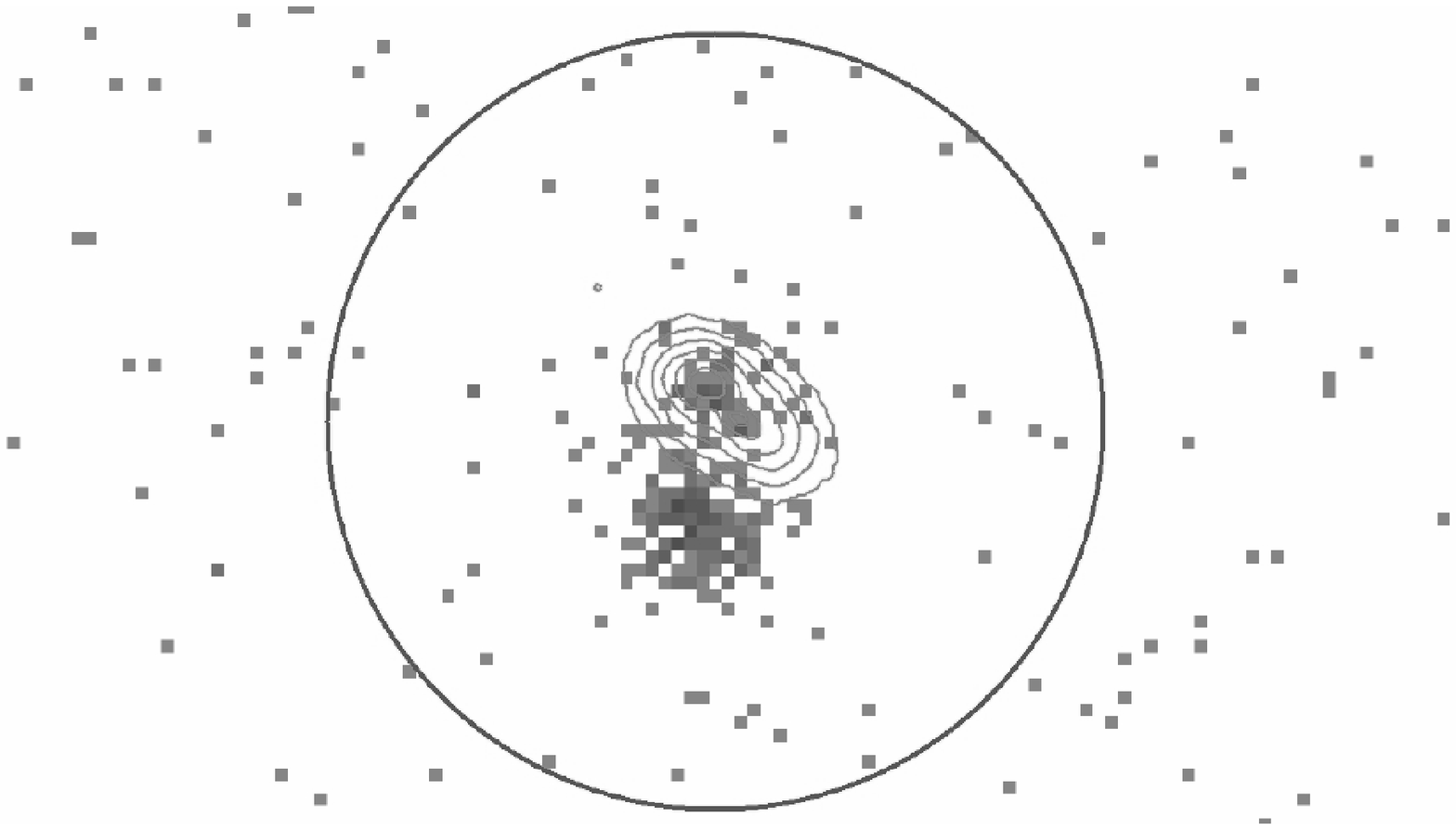}
\caption{ (left) The smoothed, summed, 148~ks exposure ACIS-I image 
centered on M31*.  The circle is the $5''$ Bondi radius. 
The color coding shows the soft band in red
(0.3--1.0 keV), the medium band in green (1.0--2.5 keV) and the hard band
in blue (2.5--7.0 keV).  There is no clear change in the temperature of
the diffuse gas at or inside of the Bondi radius. The low temperature
of the super-soft source immediately to the South of \ms is apparent
by its yellowish color. (right): The longest single ACIS-S exposure of
\ms\ is this 40ks exposure (obsid 1575).  The HST/ACS F435W 
contours are overlaid and the $5''$ Bondi radius is
shown.  Pixel size has been set to $0.125''$ to match the HRC
images shown earlier.}
\end{figure}

Because spectral fitting summed ACIS data spanning many years
consisting of 29 separate observations is not straightforward, we
carried out a spectral analysis with the longest single ACIS
observation of M31*, OBSID 1575. This 40~ks ACIS-S exposure is shown in
Figure~7 (right), with Bondi radius and also the HST contours
overlaid.  Extracting the counts in a $0.4''$ radius centered at \ms\
and excluding the overlapping region centered at P1 we find 70 counts
and find a good fit to a power-law spectrum of the form $F_{\nu,x} \propto
\nu^{-\alpha_\nu,x}$ with energy index $\alpha_{\nu,x} = 0.9 \pm 0.4$ and
\nh\ consistent with Galactic value of $6 \times 10^{20}$ \cmsq .
Freezing \nh\ to be equal to the Galactic value we are able to place a
tighter constraint on $\alpha_{\nu,x} = 0.9 \pm 0.2$.

Extracting the diffuse counts within the Bondi radius, and excluding
the point sources, we find 182 counts.  Fits to a single component
model do not yield acceptable $\chi^2$, but fits to a power-law (which
is suitable to represent the undetected point sources and the small
amount of scattering from the detected point sources) and a thermal
spectrum are acceptable.  Freezing the power-law to $\alpha_{\nu,x} = 0.7$ and
\nh\ to $6 \times 10^{20}$ \cms (consistent with the Galactic value)
yields a good fit ($\chi^2 = 1.3$ with 9~dof) with kT$=0.34 \pm 0.05$.
The density within this region, assuming a $5''$~radius sphere, is
measured to be $\sim 0.1$\cm3 .

\subsection{VLA}

We observed \ms\ simultaneously with the VLA at 5 GHz during our
2004/2005 Chandra/HRC observations.  Figure~8 (right) shows the day-long
averages of the radio flux observed during these observations.  We did
not see any significant variability within the day long averages, to a
limit of 15\% on timescales of 2, 4, and 6 hours.  On other occasions M31* has
shown variability on several hour long timescales, as can be seen in
Figure~8 (left).  Here the hypothesis of a constant source can be
rejected at the 99.99\% confidence level ($\chi^2/\nu = 43.1/11$).
This high confidence level is due almost entirely to the two
measurements of zero flux in 2002 July 06, as if we disregard those
measurements the hypothesis of a constant source is acceptable at the
24\% confidence level ($\chi^2/\nu = 11.6/9$).  

\ms\ has been observed at several radio wavelengths from the VLA, but
not at the same time.  Because the source is variable determining a
spectrum from these observations is uncertain, but it is all the
current data will allow.  The mean flux densities observed at the VLA
are $\sim 30$ $\mu$Jy at 8.4 GHz (measured on 1990 July, 1992
November, 1994 July, 1995 July, 1996 January), $\sim 50$ $\mu$Jy at
4.9 GHz (measured on 2002 July \& August, 2003 June, July \& August,
and during our monitoring simultaneous with Chandra in 2004 December,
2005 January \& February) and between 100 and 140 $\mu$Jy at 1.4 GHz
(measured on 1981 August, 1986 August \& September).  The beam size
varies from $0.24''$ to $0.4''$ to $1.4''$, respectively at these
frequencies, so one must also beware that if there is any diffuse
emission it would elevate the fluxes at the lower frequencies, however
the available images do not show any such emission.  Taken at face
value these flux densities indicate a slope $\alpha_{\nu,r} = 0.8$, where
$F_{\nu,r} \propto \nu^{-\alpha_{\nu,r}}$.

\begin{figure}[h]
\plottwo{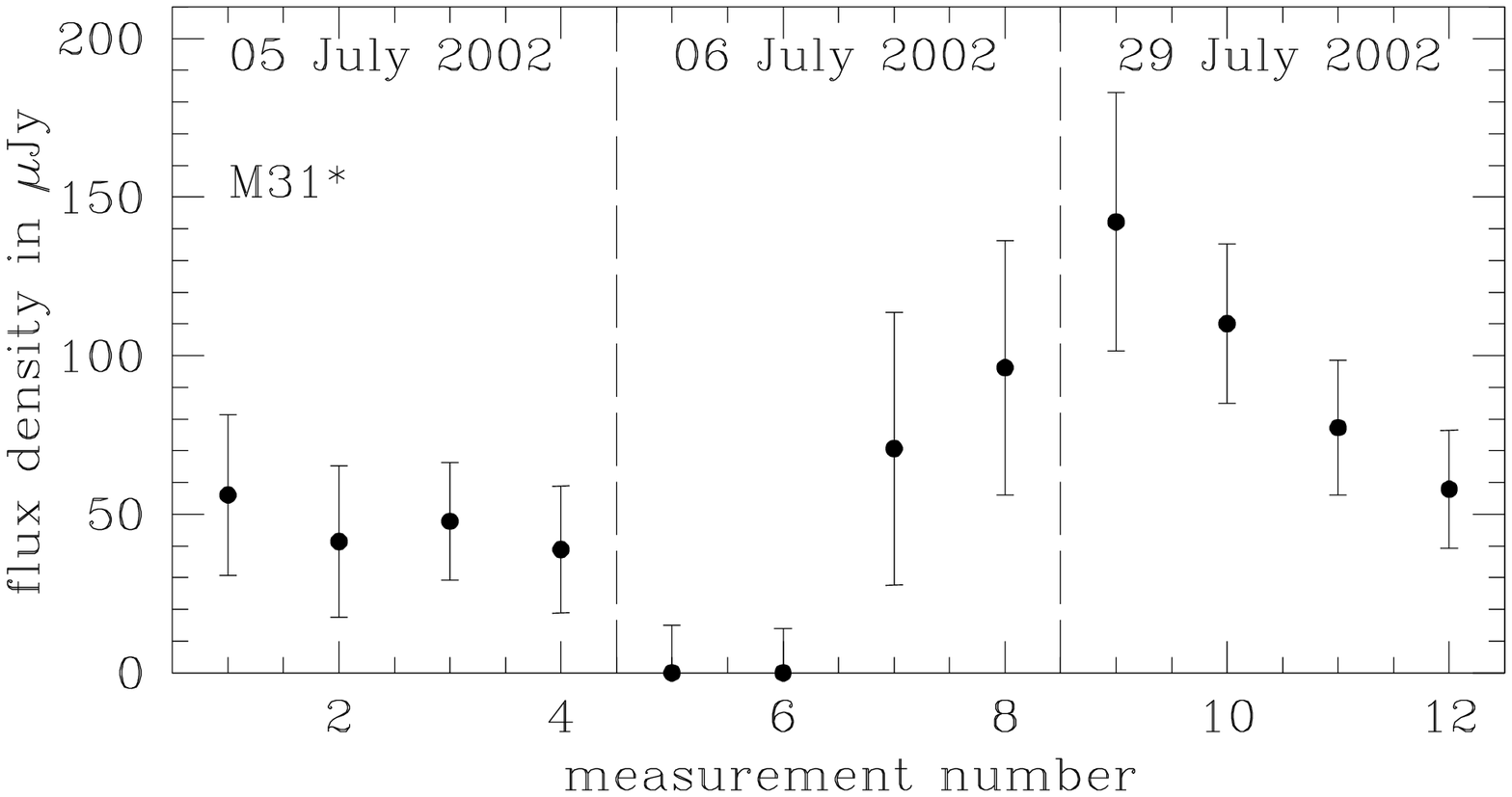}{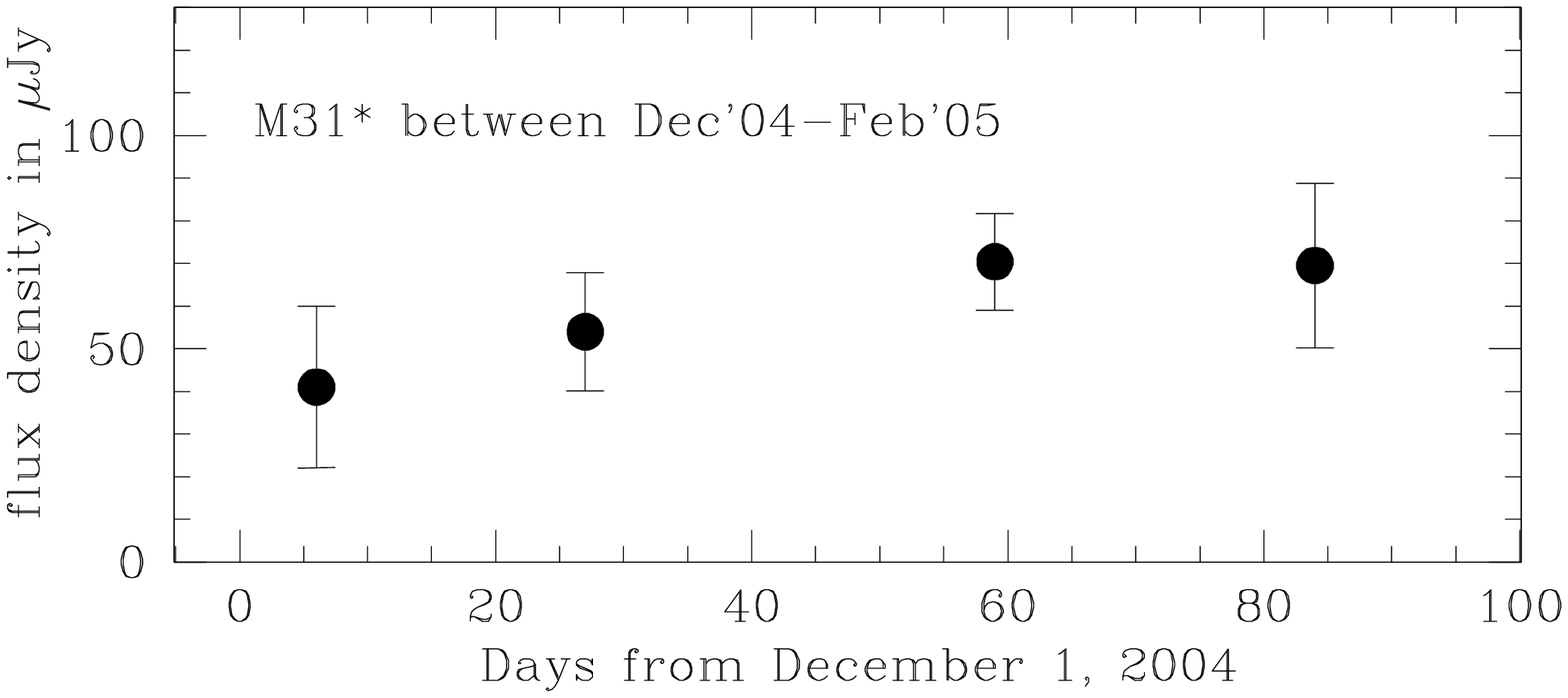}
\caption{(left) The 2-hour averaged radio variability of the nuclear
source \ms\ in three 8-hour 6 cm VLA B-configuration observations
during 2002 (Sjouwerman et al. 2005). During the first half of the
second observation \ms\ was not detected, while it was detected during
the second half, implying significant variability on a few hour long
time scale. 
(right) The day-long VLA averages at 5 GHz from our
simultaneous radio/X-ray observations during 2004/2005. 
Variability on the few hour long time scale
seen in 2002 is not evident in any of these 4
individual observations, suggesting that \ms may have been in a
different (non-variable?) state during 2004/2005.
}\label{radio}
\end{figure}

\section{Discussion}

\subsection{A Random Superposition?}

While we have detected X-ray emission at a location consistent with
\ms\, which we argue does indeed originate from \ms, we must also ask
what the chances are of a random X-ray binary within the nuclear region
of M31 being located at the position of P3=\ms ?  In order to answer
this question one needs to estimate the density of sources in the
nuclear region.  As the density increases with decreasing radius one
needs to assume some fiducial radius within which to compute the
density.

We take the Bondi radius as this fiducial radius.  We then count the
number of sources within this radius both in M31, and also in our
Galaxy under the assumption that the density of bright nuclear sources
is similar.  Within the Galaxy, the number of $>10^{36}$~\ergs\/ at the
\ms\ Bondi radius if it was transferred to the Galaxy (ie, within $100
\times 5''$) is two (Wijnands et al 2008).  Given this number, the
probability of a random source being within $0.2''$ (the
registration/centroiding error) of \ms\ is 0.3\%.  If we count sources
$>10^{36}$~\ergs\/ within the Bondi radius at M31 itself, we find 4 (or
5, depending upon the epoch) and a probability of 0.6\%.  We note that
these numbers are conservative, because if we increase the fiducial radius by
3.2x and therefore the surface area by 10x, the average source
density and therefore probability decreases by 5x. 

While the odds of an interloper are small, we note that 
there is nothing in the X-ray spectrum of light curve of the source at
P3 to distinguish it from an interloping X-ray binary.  Our
identification of this source as \ms\ is based solely on positional
coincidence.  The variations in luminosity that we have found of $0.6-20
\times 10^{36}$~\ergs (0.1-7.0 keV) are consistent with the upper limit
of $1.2 \times 10^{36}$~\ergs\ set by \cite{li.2009}, particularly
considering that \ms\ was relatively faint in the 2004/2005 observations.  

As well as confirming our tentative detection of \ms\
\citep{garcia.2005}, we have detected variable emission consistent with
a location in P1.  The emission could be due to a single low mass
X-ray binary embedded in P1.

\subsection{Bondi  Rate}

We have previously estimated the Bondi accretion rate
\citep{garcia.2005}.  Herein we update that estimation using the same
methods.  We measure the temperature of the diffuse gas within $5''$
of \ms\ to be $0.34 \pm 0.05$ keV, consistent with what has been
previously measured in the inner $\sim 1'$ \citep{taka.m31,
  li.wang.2007, bogdan.gilfanov.diffuse}.  This temperature and the
new mass estimate lead to a Bondi radius of $5.2''$ at 780~kpc.  The
density in the inner $1''$ has been estimated at 0.06 \cm3
\citep{li.wang.2007} and 0.1~\cm3 \citep{taka.m31, anil.2001}.  Herein
we find a density of 0.1~\cm3 within the Bondi radius, leading to a
Bondi accretion rate of $7 \times 10^{-5}$~\msolyear\ and a Bondi
luminosity of $4 \times 10^{41}$~\ergs .  Given the observed
luminosity of $\sim 2 \times 10^{36}$~\ergs\ this leads to an under
luminosity of $5 \times 10^{-6}$\, as shown in Figure~1.

While the Bondi rate is the standard with which to
compare SMBHs and to search for evidence of RIAFs, it assumes
that the gas is stationary at the Bondi radius.  Winds from stars,
supernovas, or other sources may modify the Bondi rate if they are
at high enough velocity \citep{melia.1992}. 
\cite{bogdan.gilfanov.diffuse} suggest that SNR drive a bipolar wind
out of the plane of M31 with a speed of $\sim 60$\kms .  This wind
is seen at large distances from \ms , but if it extends to, or
originates in, the nuclear region of M31 it still should not 
greatly effect the Bondi flow because its velocity is a factor of five 
lower than the sound speed of $\sim 300$\kms .
On the other hand, if the bipolar flow centers on \ms\ then there
may be some connection with outflows or jets originating from the RIAF.

\subsection{Variability}

In the X-ray, the most  rapid variability which is clearly detected is
a factor  of 3  on a  10~day timescale.  This  is consistent  with the
orbital timescale  at $\sim 100$ Schwarzchild radii  ($r_S = 2GM/c^2$)
from \ms  .  X-ray flares are  seen in \sgra\ with  durations of hours
and   amplitudes  of   a  factor   of   $\sim$10  \citep{bag.nat.2001,
  marrone.2008}.  Given  that we would  expect the time scale  of such
flares to  scale with mass,  similar flares in  \ms\ would occur  on a
$\sim 40$~hour  or a few day  long time scales.  Thus  this most rapid
variability we  see from \ms\  might plausibly be associated  with the
flares seen in \sgra .

Figure~6 shows an (unsuccessful) 
attempt to detect X-ray variability on even shorter
time scales.  If we had detected variability on very short time scales
it might be taken as evidence that the X-ray emission was not from
\ms , but from an interloping X-ray binary.  However, we caution here
that the $2 r_S$ light crossing time for \ms\ is $\sim 1/3$ hour, so
one would have to detect variations on time scales shorter than this
in order to argue against an origin in \ms .

In the radio, we detect more rapid variability than in the X-ray.
In particular, during the first half of an 8~hour observation on 2002
July 06 \ms\ is undetected and during the second half it is clearly
detected.  This implies variability of at least a factor of 2 on
several hour time scales.  \sgra\ typically shows only 10\% to 40\% 
variability in the radio on time scales as short as a day 
\cite{radio.var.1982, radio.var.2006}.  Given this, the variations we
see on 06 July 2002 are unusual.

The spectral slope of the emission from M31* appears to be the same in
the radio and X-ray frequency ranges, but the X-ray flux is far above
the extension of the radio spectrum into the X-ray range.  As is the
case with \sgra , this energy distribution is consistent with a
synchrotron source for the radio emission and Compton up-scattering of
the radio photons by the relativistic electrons into the X-ray range
in synchrotron self Compton or SSC process
\citep{falcke.markoff,liu.melia,hornstein2007}.  If this is indeed the
emission mechanism, then the short timescale radio variability should
be echoed in the X-ray flux.  Despite the fact that we did not detect
X-ray variability on this short timescale during the observations
presented herein, the X-ray counting rate when \ms\ is at its
brightest is sufficient to suggest that future observations may detect
this.

\subsection{Accretion Flow}

\sgra\ appears to be a slightly resolved (non point) source in the
Chandra images with a luminosity of $\sim 10^{34}$~\ergs.  This implies
that the accretion flow itself is bright in X-rays.  In contrast, in
\ms\ there is no discernible difference in the X-ray emission within
the Bondi radius.  It appears consistent with that from the diffuse
gas in the larger surrounding area.  Unfortunately, it is unclear
whether we would even detect the
accretion flow around \sgra\/ if it was at the distance of \ms .  The
fractional Eddington luminosity of an ADAF flow scales with fractional
Eddington mass accretion rate (see for example Figure~7 of
\cite{narayan.jeff.2008}.  \ms\ and \sgra\ are accreting at
approximately the same Eddington scaled rate, so the \sgra\ flow would
be be $\sim 50$x brighter around \ms\ due to its higher mass, or $\sim
5 \times 10^{35}$~\ergs .  This is on the order of the upper limit
to excess diffuse emission due to the Bondi flow that we derived above. 
However, it is important to note that any prediction based on 
a \sgra\/-like ADAF is uncertain by at least an
order of magnitude because the $0.1 - 1.0$~keV X-ray luminosity 
of the \sgra\/ flow is hidden behind $10^{23}$\cmsq\/ of absorption.

If we accept that the A-stars in P3 are from a recent star formation
episode 
then age of the P3 star cluster may be $\sim 200$~Myr
\citep{Bender.2005}, and the mass required to produce this cluster
during a single star formation burst may be between a few
$10^4$~\msol\ and $10^6$~\msol\ \citep{Bender.2005, chang.2007}
respectively.  This implies there was a source providing this mass at
a rate of $\sim 10^{-4}\ {\rm to}\ 10^{-2}$~\msol~yr$^{-1}$.   We note that the
Bondi accretion rate is comparable to the lower range of the rate
needed to form the star cluster, suggesting that the SMBH itself may 
influence the rate of star formation in its immediate surroundings
(see also \cite{fatuzzo.melia} concerning this issue with regard to
the Galactic center).
Of course, suggesting that Bondi accretion is important begs the
question of where the gas that is being accreted came from in the
first place.  We also note the co-incidence between the suggested age
of the P3 star cluster (200 Myr) and the time of the last crossing of
M32 through the nuclear region of M31 \cite{block.2006}, suggesting that this
crossing may have provided the trigger for the star formation event which 
formed the UV bright cluster at P3.

\section{Conclusions}

We have detected X-ray emission from both optical nuclei of M31 (P1
and P3) and this variable emission is consistent with two point
sources. Having resolved M31$^*$ from the surrounding point and
diffuse X-ray sources we can now investigate the accretion properties
of this nearby SMBH.  The presence of a hot and truly diffuse emission
component in the core of M31 was first noted in {\it Einstein}\/
observations \citep{tf.m31} and later confirmed with {\it ROSAT}\/
\citep{fap.rosat.m31}, { XMM-Newton}~\citep{shirey.xmm}, and {\it
  Chandra}\/ \citep{anil.diffuse} observations.  Any of this gas
within the Bondi radius of the SMBH will accrete and possibly generate
accretion luminosity.  In order to compute the Bondi accretion rate we
use the X-ray observations to estimate the temperature and density of
this gas.  The resulting rate accretion rate would produce
a luminosity of $4 \times 10^{41}$~\ergs if the gas radiated with the
canonical $\sim 10$\% efficiency.  Given the observed luminosity of 
 $\sim 2 \times 10^{36}$~\ergs , \ms\ is 
one of the most under-luminous SMBHs known. 

 Because M31* has the most highly resolved Bondi flow of any SMBH, a
 long X-ray observation could determine the applicability of ADAF,
 CDAF, ADIOS, or other models to quiescent SMBH accretion.  This in
 turn will tell us the form that black hole accretion takes for the
 vast majority of cosmic time.  For example, a 400~ks ACIS-S
 observation would yield 2500 counts within the Bondi radius (nearly
 10 per pixel), sufficient to divide the region into 5 annuli and
 determine accurate temperatures in 5 radial rings.  This would be
 sufficient to determine the run of temperature with radius in the
 RIAF. The various RIAF models all predict $T~(1/r)^\alpha$, so these
 data could determine $\alpha$ and therefore the structure of the
 RIAF.

\section{Acknowledgments}

This work was supported in part by \chandra\ Grant GO-6088A and
Chandra X-ray Center Contract NAS8-03060.

\facility{Chandra}
\facility{HST(ACS)}

\vspace{-0.5cm}

\bibliographystyle{apj}
\bibliography{apjmnemonic,references}

\begin{thebibliography}{63}
\expandafter\ifx\csname natexlab\endcsname\relax\def\natexlab#1{#1}\fi

\bibitem[{{Baganoff} {et~al.}(2001){Baganoff}, {Bautz}, {Brandt}, {Chartas},
  {Feigelson}, {Garmire}, {Maeda}, {Morris}, {Ricker}, {Townsley}, \&
  {Walter}}]{bag.nat.2001}
{Baganoff}, F.~K., {Bautz}, M.~W., {Brandt}, W.~N., {Chartas}, G., {Feigelson},
  E.~D., {Garmire}, G.~P., {Maeda}, Y., {Morris}, M., {Ricker}, G.~R.,
  {Townsley}, L.~K., \& {Walter}, F. 2001, \nat, 413, 45

\bibitem[{{Baganoff} {et~al.}(2003){Baganoff}, {Maeda}, {Morris}, {Bautz},
  {Brandt}, {Cui}, {Doty}, {Feigelson}, {Garmire}, {Pravdo}, {Ricker}, \&
  {Townsley}}]{baganoff.2003}
{Baganoff}, F.~K., {Maeda}, Y., {Morris}, M., {Bautz}, M.~W., {Brandt}, W.~N.,
  {Cui}, W., {Doty}, J.~P., {Feigelson}, E.~D., {Garmire}, G.~P., {Pravdo},
  S.~H., {Ricker}, G.~R., \& {Townsley}, L.~K. 2003, \apj, 591, 891

\bibitem[{{Barmby} {et~al.}(2006){Barmby}, {Ashby}, {Bianchi}, {Engelbracht},
  {Gehrz}, {Gordon}, {Hinz}, {Huchra}, {Humphreys}, {Pahre},
  {P{\'e}rez-Gonz{\'a}lez}, {Polomski}, {Rieke}, {Thilker}, {Willner}, \&
  {Woodward}}]{barmby.2006}
{Barmby}, P., {Ashby}, M.~L.~N., {Bianchi}, L., {Engelbracht}, C.~W., {Gehrz},
  R.~D., {Gordon}, K.~D., {Hinz}, J.~L., {Huchra}, J.~P., {Humphreys}, R.~M.,
  {Pahre}, M.~A., {P{\'e}rez-Gonz{\'a}lez}, P.~G., {Polomski}, E.~F., {Rieke},
  G.~H., {Thilker}, D.~A., {Willner}, S.~P., \& {Woodward}, C.~E. 2006, \apjl,
  650, L45

\bibitem[{{Barmby} \& {Huchra}(2001)}]{barmby.huchra.2001}
{Barmby}, P. \& {Huchra}, J.~P. 2001, \aj, 122, 2458

\bibitem[{{Beckerman} {et~al.}(2004){Beckerman}, {Aldcroft}, {Gaetz}, {Jerius},
  {Nguyen}, \& {Tibbetts}}]{aldcroft.2004}
{Beckerman}, E., {Aldcroft}, T., {Gaetz}, T.~J., {Jerius}, D.~H., {Nguyen}, D.,
  \& {Tibbetts}, M. 2004, in Presented at the Society of Photo-Optical
  Instrumentation Engineers (SPIE) Conference, Vol. 5165, Society of
  Photo-Optical Instrumentation Engineers (SPIE) Conference Series, ed. K.~A.
  {Flanagan} \& O.~H.~W. {Siegmund}, 445--456

\bibitem[{{Bender} {et~al.}(2005){Bender}, {Kormendy}, {Bower}, {Green},
  {Thomas}, {Danks}, {Gull}, {Hutchings}, {Joseph}, {Kaiser}, {Lauer},
  {Nelson}, {Richstone}, {Weistrop}, \& {Woodgate}}]{Bender.2005}
{Bender}, R., {Kormendy}, J., {Bower}, G., {Green}, R., {Thomas}, J., {Danks},
  A.~C., {Gull}, T., {Hutchings}, J.~B., {Joseph}, C.~L., {Kaiser}, M.~E.,
  {Lauer}, T.~R., {Nelson}, C.~H., {Richstone}, D., {Weistrop}, D., \&
  {Woodgate}, B. 2005, \apj, 631, 280

\bibitem[{{Blandford} \& {Begelman}(1999)}]{bb.adios}
{Blandford}, R.~D. \& {Begelman}, M.~C. 1999, \mnras, 303, L1

\bibitem[{{Block} {et~al.}(2006){Block}, {Bournaud}, {Combes}, {Groess},
  {Barmby}, {Ashby}, {Fazio}, {Pahre}, \& {Willner}}]{block.2006}
{Block}, D.~L., {Bournaud}, F., {Combes}, F., {Groess}, R., {Barmby}, P.,
  {Ashby}, M.~L.~N., {Fazio}, G.~G., {Pahre}, M.~A., \& {Willner}, S.~P. 2006,
  \nat, 443, 832

\bibitem[{{Bogdan} \& {Gilfanov}(2008)}]{bogdan.gilfanov.diffuse}
{Bogdan}, A. \& {Gilfanov}, M. 2008, ArXiv e-prints, 803

\bibitem[{{Brown} \& {Lo}(1982)}]{radio.var.1982}
{Brown}, R.~L. \& {Lo}, K.~Y. 1982, \apj, 253, 108

\bibitem[{{Chang} {et~al.}(2007){Chang}, {Murray-Clay}, {Chiang}, \&
  {Quataert}}]{chang.2007}
{Chang}, P., {Murray-Clay}, R., {Chiang}, E., \& {Quataert}, E. 2007, \apj,
  668, 236

\bibitem[{{Crane} {et~al.}(1993){Crane}, {Dickel}, \& {Cowan}}]{crane.93}
{Crane}, P.~C., {Dickel}, J.~R., \& {Cowan}, J.~J. 1993, \apjl, 411, L107+

\bibitem[{{Demarque} \& {Virani}(2007)}]{demarque.virani}
{Demarque}, P. \& {Virani}, S. 2007, \aap, 461, 651

\bibitem[{{Di Stefano} {et~al.}(2004){Di Stefano}, {Kong}, {Greiner},
  {Primini}, {Garcia}, {Barmby}, {Massey}, {Hodge}, {Williams}, {Murray},
  {Curry}, \& {Russo}}]{distefano.2004}
{Di Stefano}, R., {Kong}, A.~K.~H., {Greiner}, J., {Primini}, F.~A., {Garcia},
  M.~R., {Barmby}, P., {Massey}, P., {Hodge}, P.~W., {Williams}, B.~F.,
  {Murray}, S.~S., {Curry}, S., \& {Russo}, T.~A. 2004, \apj, 610, 247

\bibitem[{{Dosaj} {et~al.}(2002){Dosaj}, {Garcia}, {Forman}, {Jones}, {Kong},
  {di Stefano}, {Primini}, \& {Murray}}]{anil.diffuse}
{Dosaj}, A., {Garcia}, M., {Forman}, W., {Jones}, C., {Kong}, A., {di Stefano},
  R., {Primini}, F., \& {Murray}, S. 2002, in ASP Conf. Ser. 262: The High
  Energy Universe at Sharp Focus: Chandra Science, 147--+

\bibitem[{{Dosaj} {et~al.}(2001){Dosaj}, {Garcia}, {Forman}, {Jones}, {Kong},
  {Primini}, {Di Stefano}, \& {Murray}}]{anil.2001}
{Dosaj}, A., {Garcia}, M.~G., {Forman}, W.~R., {Jones}, C., {Kong}, A.,
  {Primini}, F.~A., {Di Stefano}, R., \& {Murray}, S.~S. 2001, in Bulletin of
  the American Astronomical Society, Vol.~33, Bulletin of the American
  Astronomical Society, 1369--+

\bibitem[{{Edmonds} {et~al.}(2003){Edmonds}, {Gilliland}, {Heinke}, \&
  {Grindlay}}]{edmonds.2003}
{Edmonds}, P.~D., {Gilliland}, R.~L., {Heinke}, C.~O., \& {Grindlay}, J.~E.
  2003, \apj, 596, 1177

\bibitem[{{Falcke} \& {Markoff}(2000)}]{falcke.markoff}
{Falcke}, H. \& {Markoff}, S. 2000, \aap, 362, 113

\bibitem[{{Fatuzzo} \& {Melia}(2009)}]{fatuzzo.melia}
{Fatuzzo}, M. \& {Melia}, F. 2009, \pasp, 121, 585

\bibitem[{{Garcia} {et~al.}(2001){Garcia}, {Kong}, {Primini}, {Barmby}, {Di
  Stefano}, {McClintock}, \& {Murray}}]{garcia.m31*not.2001}
{Garcia}, M.~R., {Kong}, A., {Primini}, F.~A., {Barmby}, P., {Di Stefano}, R.,
  {McClintock}, J.~E., \& {Murray}, S.~S. 2001, in Two Years of Science with
  Chandra, Abstracts from the Symposium held in Washington, DC, 5-7 September,
  2001.

\bibitem[{{Garcia} {et~al.}(2000){Garcia}, {Murray}, {Primini}, {Forman},
  {McClintock}, \& {Jones}}]{garcia.m31*.2000}
{Garcia}, M.~R., {Murray}, S.~S., {Primini}, F.~A., {Forman}, W.~R.,
  {McClintock}, J.~E., \& {Jones}, C. 2000, \apjl, 537, L23

\bibitem[{{Garcia} {et~al.}(2005){Garcia}, {Williams}, {Yuan}, {Kong},
  {Primini}, {Barmby}, {Kaaret}, \& {Murray}}]{garcia.2005}
{Garcia}, M.~R., {Williams}, B.~F., {Yuan}, F., {Kong}, A.~K.~H., {Primini},
  F.~A., {Barmby}, P., {Kaaret}, P., \& {Murray}, S.~S. 2005, \apj, 632, 1042

\bibitem[{{Ghez} {et~al.}(2003{\natexlab{a}}){Ghez}, {Duch{\^e}ne}, {Matthews},
  {Hornstein}, {Tanner}, {Larkin}, {Morris}, {Becklin}, {Salim}, {Kremenek},
  {Thompson}, {Soifer}, {Neugebauer}, \& {McLean}}]{ghez.youth.2003}
{Ghez}, A.~M., {Duch{\^e}ne}, G., {Matthews}, K., {Hornstein}, S.~D., {Tanner},
  A., {Larkin}, J., {Morris}, M., {Becklin}, E.~E., {Salim}, S., {Kremenek},
  T., {Thompson}, D., {Soifer}, B.~T., {Neugebauer}, G., \& {McLean}, I.
  2003{\natexlab{a}}, \apjl, 586, L127

\bibitem[{{Ghez} {et~al.}(2003{\natexlab{b}}){Ghez}, {Duch{\^e}ne}, {Matthews},
  {Hornstein}, {Tanner}, {Larkin}, {Morris}, {Becklin}, {Salim}, {Kremenek},
  {Thompson}, {Soifer}, {Neugebauer}, \& {McLean}}]{ghez.2003}
---. 2003{\natexlab{b}}, \apjl, 586, L127

\bibitem[{{Gordon} {et~al.}(2006){Gordon}, {Bailin}, {Engelbracht}, {Rieke},
  {Misselt}, {Latter}, {Young}, {Ashby}, {Barmby}, {Gibson}, {Hines}, {Hinz},
  {Krause}, {Levine}, {Marleau}, {Noriega-Crespo}, {Stolovy}, {Thilker}, \&
  {Werner}}]{gordon.2006}
{Gordon}, K.~D., {Bailin}, J., {Engelbracht}, C.~W., {Rieke}, G.~H., {Misselt},
  K.~A., {Latter}, W.~B., {Young}, E.~T., {Ashby}, M.~L.~N., {Barmby}, P.,
  {Gibson}, B.~K., {Hines}, D.~C., {Hinz}, J., {Krause}, O., {Levine}, D.~A.,
  {Marleau}, F.~R., {Noriega-Crespo}, A., {Stolovy}, S., {Thilker}, D.~A., \&
  {Werner}, M.~W. 2006, \apjl, 638, L87

\bibitem[{{Hawley} \& {Balbus}(2002)}]{hb.adios}
{Hawley}, J.~F. \& {Balbus}, S.~A. 2002, \apj, 573, 738

\bibitem[{{Henze} {et~al.}(2009){Henze}, {Pietsch}, {Sala}, {Della Valle},
  {Hernanz}, {Greiner}, {Burwitz}, {Freyberg}, {Haberl}, {Hartmann}, {Milne},
  \& {Williams}}]{henze.2009}
{Henze}, M., {Pietsch}, W., {Sala}, G., {Della Valle}, M., {Hernanz}, M.,
  {Greiner}, J., {Burwitz}, V., {Freyberg}, M.~J., {Haberl}, F., {Hartmann},
  D.~H., {Milne}, P., \& {Williams}, G.~G. 2009, \aap, 498, L13

\bibitem[{{Hornstein} {et~al.}(2007){Hornstein}, {Matthews}, {Ghez}, {Lu},
  {Morris}, {Becklin}, {Rafelski}, \& {Baganoff}}]{hornstein2007}
{Hornstein}, S.~D., {Matthews}, K., {Ghez}, A.~M., {Lu}, J.~R., {Morris}, M.,
  {Becklin}, E.~E., {Rafelski}, M., \& {Baganoff}, F.~K. 2007, \apj, 667, 900

\bibitem[{{Igumenshchev} {et~al.}(2000){Igumenshchev}, {Abramowicz}, \&
  {Narayan}}]{cdaf.Ig}
{Igumenshchev}, I.~V., {Abramowicz}, M.~A., \& {Narayan}, R. 2000, \apjl, 537,
  L27

\bibitem[{{Kaaret}(2002)}]{kaaret.hrc}
{Kaaret}, P. 2002, \apj, 578, 114

\bibitem[{{Kong} {et~al.}(2003){Kong}, {DiStefano}, {Garcia}, \&
  {Greiner}}]{kong.2003}
{Kong}, A.~K.~H., {DiStefano}, R., {Garcia}, M.~R., \& {Greiner}, J. 2003,
  \apj, 585, 298

\bibitem[{{Kong} {et~al.}(2002{\natexlab{a}}){Kong}, {Garcia}, {Primini}, \&
  {Murray}}]{kong.snr1}
{Kong}, A.~K.~H., {Garcia}, M.~R., {Primini}, F.~A., \& {Murray}, S.~S.
  2002{\natexlab{a}}, \apjl, 580, L125

\bibitem[{{Kong} {et~al.}(2002{\natexlab{b}}){Kong}, {Garcia}, {Primini},
  {Murray}, {Di Stefano}, \& {McClintock}}]{kong.m31.l1}
{Kong}, A.~K.~H., {Garcia}, M.~R., {Primini}, F.~A., {Murray}, S.~S., {Di
  Stefano}, R., \& {McClintock}, J.~E. 2002{\natexlab{b}}, \apj, 577, 738

\bibitem[{{Kormendy} \& {Bender}(1999)}]{kormendy.bender.99}
{Kormendy}, J. \& {Bender}, R. 1999, \apj, 522, 772

\bibitem[{{Lauer} {et~al.}(1993){Lauer}, {Faber}, {Groth}, {Shaya}, {Campbell},
  {Code}, {Currie}, {Baum}, {Ewald}, {Hester}, {Holtzman}, {Kristian}, {Light},
  {Ligynds}, {O'Neil}, \& {Westphal}}]{lauer.93}
{Lauer}, T.~R., {Faber}, S.~M., {Groth}, E.~J., {Shaya}, E.~J., {Campbell}, B.,
  {Code}, A., {Currie}, D.~G., {Baum}, W.~A., {Ewald}, S.~P., {Hester}, J.~J.,
  {Holtzman}, J.~A., {Kristian}, J., {Light}, R.~M., {Ligynds}, C.~R.,
  {O'Neil}, E.~J., \& {Westphal}, J.~A. 1993, \aj, 106, 1436

\bibitem[{{Li} \& {Wang}(2007{\natexlab{a}})}]{li.wang.2007}
{Li}, Z. \& {Wang}, Q.~D. 2007{\natexlab{a}}, \apjl, 668, L39

\bibitem[{{Li} \& {Wang}(2007{\natexlab{b}})}]{li.wang.diffuse}
---. 2007{\natexlab{b}}, \apjl, 668, L39

\bibitem[{{Li} {et~al.}(2009){Li}, {Wang}, \& {Wakker}}]{li.2009}
{Li}, Z., {Wang}, Q.~D., \& {Wakker}, B.~P. 2009, \mnras, 397, 148

\bibitem[{{Liu} \& {Melia}(2001)}]{liu.melia}
{Liu}, S. \& {Melia}, F. 2001, \apjl, 561, L77

\bibitem[{{Macquart} \& {Bower}(2006)}]{radio.var.2006}
{Macquart}, J.-P. \& {Bower}, G.~C. 2006, \apj, 641, 302

\bibitem[{{Marrone} {et~al.}(2008){Marrone}, {Baganoff}, {Morris}, {Moran},
  {Ghez}, {Hornstein}, {Dowell}, {Mu{\~n}oz}, {Bautz}, {Ricker}, {Brandt},
  {Garmire}, {Lu}, {Matthews}, {Zhao}, {Rao}, \& {Bower}}]{marrone.2008}
{Marrone}, D.~P., {Baganoff}, F.~K., {Morris}, M.~R., {Moran}, J.~M., {Ghez},
  A.~M., {Hornstein}, S.~D., {Dowell}, C.~D., {Mu{\~n}oz}, D.~J., {Bautz},
  M.~W., {Ricker}, G.~R., {Brandt}, W.~N., {Garmire}, G.~P., {Lu}, J.~R.,
  {Matthews}, K., {Zhao}, J.-H., {Rao}, R., \& {Bower}, G.~C. 2008, \apj, 682,
  373

\bibitem[{{Martins} {et~al.}(2008){Martins}, {Gillessen}, {Eisenhauer},
  {Genzel}, {Ott}, \& {Trippe}}]{martins.2008}
{Martins}, F., {Gillessen}, S., {Eisenhauer}, F., {Genzel}, R., {Ott}, T., \&
  {Trippe}, S. 2008, \apjl, 672, L119

\bibitem[{{Massey} {et~al.}(2006){Massey}, {Olsen}, {Hodge}, {Strong},
  {Jacoby}, {Schlingman}, \& {Smith}}]{lgs}
{Massey}, P., {Olsen}, K.~A.~G., {Hodge}, P.~W., {Strong}, S.~B., {Jacoby},
  G.~H., {Schlingman}, W., \& {Smith}, R.~C. 2006, \aj, 131, 2478

\bibitem[{{Melia}(1992{\natexlab{a}})}]{melia.1992}
{Melia}, F. 1992{\natexlab{a}}, \apjl, 387, L25

\bibitem[{{Melia}(1992{\natexlab{b}})}]{melia92}
---. 1992{\natexlab{b}}, \apjl, 398, L95

\bibitem[{{Narayan} {et~al.}(2000){Narayan}, {Igumenshchev}, \&
  {Abramowicz}}]{cdaf.narayan}
{Narayan}, R., {Igumenshchev}, I.~V., \& {Abramowicz}, M.~A. 2000, \apj, 539,
  798

\bibitem[{{Narayan} \& {McClintock}(2008)}]{narayan.jeff.2008}
{Narayan}, R. \& {McClintock}, J.~E. 2008, New Astronomy Review, 51, 733

\bibitem[{{Narayan} \& {Yi}(1994)}]{narayan.yi.94}
{Narayan}, R. \& {Yi}, I. 1994, \apjl, 428, L13

\bibitem[{{Narayan} \& {Yi}(1995)}]{narayan.wind.95}
---. 1995, \apj, 444, 231

\bibitem[{{Pietsch} {et~al.}(2007){Pietsch}, {Haberl}, {Sala}, {Stiele},
  {Hornoch}, {Riffeser}, {Fliri}, {Bender}, {B{\"u}hler}, {Burwitz}, {Greiner},
  \& {Seitz}}]{pietsch.2007}
{Pietsch}, W., {Haberl}, F., {Sala}, G., {Stiele}, H., {Hornoch}, K.,
  {Riffeser}, A., {Fliri}, J., {Bender}, R., {B{\"u}hler}, S., {Burwitz}, V.,
  {Greiner}, J., \& {Seitz}, S. 2007, \aap, 465, 375

\bibitem[{{Prestwich} {et~al.}(2003){Prestwich}, {Irwin}, {Kilgard}, {Krauss},
  {Zezas}, {Primini}, {Kaaret}, \& {Boroson}}]{andyp.cc}
{Prestwich}, A.~H., {Irwin}, J.~A., {Kilgard}, R.~E., {Krauss}, M.~I., {Zezas},
  A., {Primini}, F., {Kaaret}, P., \& {Boroson}, B. 2003, \apj, 595, 719

\bibitem[{{Primini} {et~al.}(1993){Primini}, {Forman}, \&
  {Jones}}]{fap.rosat.m31}
{Primini}, F.~A., {Forman}, W., \& {Jones}, C. 1993, \apj, 410, 615

\bibitem[{{Quataert} \& {Gruzinov}(2000)}]{cdaf.quat}
{Quataert}, E. \& {Gruzinov}, A. 2000, \apj, 539, 809

\bibitem[{{Shirey} {et~al.}(2001{\natexlab{a}}){Shirey}, {Soria}, {Borozdin},
  {Osborne}, {Tiengo}, {Guainazzi}, {Hayter}, {La Palombara}, {Mason},
  {Molendi}, {Paerels}, {Pietsch}, {Priedhorsky}, {Read}, {Watson}, \&
  {West}}]{shirey.xmm}
{Shirey}, R., {Soria}, R., {Borozdin}, K., {Osborne}, J.~P., {Tiengo}, A.,
  {Guainazzi}, M., {Hayter}, C., {La Palombara}, N., {Mason}, K., {Molendi},
  S., {Paerels}, F., {Pietsch}, W., {Priedhorsky}, W., {Read}, A.~M., {Watson},
  M.~G., \& {West}, R.~G. 2001{\natexlab{a}}, \aap, 365, L195

\bibitem[{{Shirey} {et~al.}(2001{\natexlab{b}}){Shirey}, {Soria}, {Borozdin},
  {Osborne}, {Tiengo}, {Guainazzi}, {Hayter}, {La Palombara}, {Mason},
  {Molendi}, {Paerels}, {Pietsch}, {Priedhorsky}, {Read}, {Watson}, \&
  {West}}]{shirey.2001}
---. 2001{\natexlab{b}}, \aap, 365, L195

\bibitem[{{Shvartsman}(1971)}]{shvartsman.1971}
{Shvartsman}, V.~F. 1971, Soviet Astronomy, 15, 377

\bibitem[{{Sjouwerman} {et~al.}(2005){Sjouwerman}, {Kong}, {Garcia}, {Dickel},
  {Williams}, {Johnson}, {Primini}, \& {Goss}}]{lorant.2005}
{Sjouwerman}, L.~O., {Kong}, A.~K.~H., {Garcia}, M.~R., {Dickel}, J.~R.,
  {Williams}, B.~F., {Johnson}, K.~E., {Primini}, F.~A., \& {Goss}, W.~M. 2005,
  in X-Ray and Radio Connections (eds. L.O. Sjouwerman and K.K Dyer) Published
  electronically by NRAO, http://www.aoc.nrao.edu/events/xraydio Held 3-6
  February 2004 in Santa Fe, New Mexico, USA, (E4.16) 4 pages, ed. L.~O.
  {Sjouwerman} \& K.~K. {Dyer}

\bibitem[{{Takahashi} {et~al.}(2004){Takahashi}, {Okada}, {Kokubun}, \&
  {Makishima}}]{taka.m31}
{Takahashi}, H., {Okada}, Y., {Kokubun}, M., \& {Makishima}, K. 2004, \apj,
  615, 242

\bibitem[{{Tremaine}(1995)}]{tremaine.1995}
{Tremaine}, S. 1995, \aj, 110, 628

\bibitem[{{Trinchieri} \& {Fabbiano}(1991)}]{tf.m31}
{Trinchieri}, G. \& {Fabbiano}, G. 1991, \apj, 382, 82

\bibitem[{{Trudolyubov} {et~al.}(2005){Trudolyubov}, {Kotov}, {Priedhorsky},
  {Cordova}, \& {Mason}}]{trudy.xmm}
{Trudolyubov}, S., {Kotov}, O., {Priedhorsky}, W., {Cordova}, F., \& {Mason},
  K. 2005, \apj, 634, 314

\bibitem[{{Williams} {et~al.}(2005){Williams}, {Garcia}, {McClintock}, {Kong},
  {Primini}, \& {Murray}}]{williams.2005}
{Williams}, B.~F., {Garcia}, M.~R., {McClintock}, J.~E., {Kong}, A.~K.~H.,
  {Primini}, F.~A., \& {Murray}, S.~S. 2005, \apj, 628, 382

\bibitem[{{Williams} {et~al.}(2006){Williams}, {Naik}, {Garcia}, \&
  {Callanan}}]{williams.2006}
{Williams}, B.~F., {Naik}, S., {Garcia}, M.~R., \& {Callanan}, P.~J. 2006,
  \apj, 643, 356

\end{thebibliography}
\end{document}